\RequirePackage{fix-cm}
\documentclass[10pt]{article} %
\usepackage{amsmath}
\usepackage{amssymb}
\usepackage{amsfonts}
\usepackage[T1]{fontenc}
\usepackage{graphicx}%
\usepackage{epsfig}%
\usepackage{calc}%
\usepackage{longtable}%
\usepackage{lscape}%
\usepackage{multirow}%
\usepackage[tableposition=top]{caption}

\newcommand{\pa}[1]{\left( #1 \right)}
\newcommand{\br}[1]{\left[ #1 \right]}
\newcommand{\ac}[1]{\left\{ #1 \right\}}
\newcommand{\abs}[1]{| #1 |}

\newcommand{\RR}{\mathbb{R}}

\newcommand{\rarrow}{\rightarrow}

\newcommand{\rank}{\operatorname{rank}}

\newcommand{\arctanh}{\operatorname{arctanh}}

\newcommand{\vl}{\vec{\lambda}}

\newcommand{\del}{\partial}

\newenvironment{aleq}{\begin{equation}\begin{aligned}}{\end{aligned}\end{equation}}
\newenvironment{aleq*}{\begin{equation*}\begin{aligned}}{\end{aligned}\end{equation*}}
\newenvironment{gaeq}{\begin{equation}\begin{gathered}}{\end{gathered}\end{equation}}
\newenvironment{gaeq*}{\begin{equation*}\begin{gathered}}{\end{gathered}\end{equation*}}
\newenvironment{eqe}{\begin{equation}}{\end{equation}}

\begin{document}
\title{Fluid dynamics solutions obtained from the Riemann invariant approach.
}




\fontsize{10}{10} \selectfont
\author{\fontsize{11}{11} \selectfont A.M. Grundland,\\
              Centre de Recherches Math\'ematiques,\\
               Universit\'e du Montr\'eal,\\
               C.P. 6128, Succc. Centre-ville, Montr\'eal, (QC) H3C 3J7, Canada,\\
and D\'epartement de math\'ematiques et informatiques,\\
Universit\'e du Qu\'ebec, Trois-Rivi\`eres, (QC) G9A 5H7, Canada,\\
              grundlan@crm.umontreal.ca      
           \and
\fontsize{11}{11} \selectfont V. Lamothe,\\
              D\'epartement de Math\'ematiques et Statistique,\\
           Universit\'e de Montr\'eal,\\
C.P. 6128, Succc. Centre-ville, Montr\'eal, (QC) H3C 3J7, Canada,\\
            lamothe@crm.umontreal.ca
 }

\maketitle

\begin{abstract}
The generalized method of characteristics is used to obtain rank-2 solutions of the classical equations of hydrodynamics  in (3+1) dimensions describing the motion of a fluid medium in the presence of gravitational and Coriolis forces. We determine the necessary and sufficient conditions which guarantee the existence of solutions expressed in terms of Riemann invariants for an inhomogeneous quasilinear system of partial differential equations. The paper contains a detailed exposition of the theory of simple wave solutions and a presentation of the main tool used to study the Cauchy problem. A systematic use is made of the generalized method of characteristics in order to generate several classes of wave solutions written in terms of Riemann invariants.
\end{abstract}

\ \\[1mm]
This work is dedicated to the memory of professor Marek Burnat.\\[2mm]

\noindent\textbf{keywords: }generalized method of characteristics, Riemann invariants, multiwave solutions, fluid dynamics equations.\\[2mm]
\noindent Mathematics Subject Classification (2000): 35B06, 35F50, 35F20
\clearpage
\section{Introduction. Inhomogeneous fluid dynamics system}
The compressible flow of an ideal fluid in the presence of gravitational and Coriolis forces is governed by the Euler equations in (3+1) dimensions
\begin{aleq}\label{eq:euler}
&\rho\pa{\frac{\del \vec{v}}{\del t}+\pa{\vec{v}\cdot\nabla}\vec{v}}+\nabla p=\rho\pa{\vec{g}-\vec{\Omega}\times \vec{v}},\\
&\frac{\del \rho}{\del t}+\nabla\cdot\pa{\rho \vec{v}}=0,\\
&\pa{\frac{\del }{\del t}+\vec{v}\cdot\nabla}\frac{p}{\rho^{\kappa}}=0,
\end{aleq}%
which constitute a hyperbolic quasilinear system in four independent variables $x=(t,\vec{x})$ (namely time $t$ and the three space variables $(x^1,x^2,x^3)=\vec{x})$ and five dependent variables: $\rho$ is the density of the fluid, $p$ is the pressure of the fluid, $\vec{v}=\pa{v^1, v^2, v^3}$ is the vector field of the fluid velocity and $\kappa$ is the adiabatic exponent. The ideal fluid flow is subjected to the gravitational force $\rho\vec{g}$ and the Coriolis force $\rho \vec{\Omega}\times\vec{v}$ in a non-inertial coordinate system.
\paragraph{}This system admits two distinct families of characteristics associated with entropic and acoustic waves. These waves play an essential role from the point of view of a physical and mathematical analysis of the initial system (\ref{eq:euler}). It is convenient to choose characteristic coordinates as new independent variables instead of the Euler or Lagrange coordinates (see \textit{e.g.} \cite{Jeffrey:1976,Lighthill:1968,Mises:1958,Rozdestvenski:1983,Whitham:1974,Zakharov:1998}). The existence of Riemann invariants which remain constant along these characteristics considerably simplifies the problem of constructing and investigating wave solutions admitted by the system (\ref{eq:euler}). An advantage of the presence of Riemann invariants (\ref{eq:euler}) is that, at least in certain favorable cases, they lead to expressions for which the general integrals are given in closed form. The objective of this paper is to look for certain classes of solutions describing the propagation of waves that satisfy the Euler equations (\ref{eq:euler}). Such classes of solutions are particularly interesting from the physical point of view because they cover a wide range of nonlinear wave phenomena arising in the presence of the external forces that are observed in fluid dynamics. The methodological approach assumed in this work is based on the generalized method of characteristics initiated by M. Burnat \cite{Burnat:1969,Burnat:1969:2} and next developed by Z. Peradzynski \cite{Peradzynski:1971:planewave,Peradzynski:1985} for homogeneous nonelliptic quasilinear systems. A specific feature of that approach is an algebraization of the partial differential equations (PDEs) under consideration by representing the general integral elements as linear combinations of some special rank-1 elements associated with certain vector fields which generate characteristic curves in the spaces of independent and dependent variables, respectively. The introduction of those rank-1 elements (also called simple elements, see definition 2.1) proved to be a useful tool for constructing solutions in the case of the inhomogeneous Euler equations. These simple integral elements are in one-to-one correspondence with Riemann wave solutions (also called simple waves). By means of the Cartan theory for involutive systems, it was shown in \cite{Grundland:1974:a,Peradzynski:1971:planewave} that these elements serve as building blocks for constructing certain classes of solutions expressed in terms of Riemann invariants which can be interpreted as multiple wave superpositions of two or more single Riemann waves \cite{Burnat:1969,Burnat:1969:2,GrundlandZelazny:1983,Peradzynski:1985}.
\paragraph{}Riemann \cite{Riemann:1858} demonstrated that simple wave solutions for hydrodynamics-type systems, even with arbitrary smooth initial data, usually cannot be extended indefinitely in time but blows up after a certain finite period. The first derivative of a simple wave solution of (\ref{eq:2.1}) becomes unbounded after some finite time, $T>0$, and for times $t>T$ smooth solutions of the initial Cauchy problem do not exist. Therefore, we deal here with the phenomenon known as the gradient catastrophe. Riemann also investigated the problem of extending the simple wave solutions in some generalized sense beyond the time $t>T$ of the blow up. On the basis of the laws of conservation of mass and momentum, Riemann introduced solutions based on discontinuous functions which can be interpreted as shock waves. He demonstrated a link between the wave front velocity and parameters of the fluid state before and behind that discontinuity. The problem of the propagation and superposition of simple waves has been extensively developed by many authors (\textit{e.g.} see
\cite{GrundlandZelazny:1983,Jeffrey:1976,Lighthill:1968,Madja:1984,Mises:1958,Peradzynski:1971:planewave,Peradzynski:1985,Riemann:1858,Rozdestvenski:1967,Rozdestvenski:1983,Whitham:1974,Zakharov:1998} and references therein).
\paragraph{}In this paper we concentrate on the simplest case, namely on the propagation of a single simple wave admitted by the inhomogeneous system (\ref{eq:euler}) and show that this leads to several new classes of interesting solutions. We also solve the problem of determining the necessary and sufficient conditions on the initial data for the corresponding Cauchy problem for the inhomogeneous quasilinear system in order that the solution evolve as a Riemann wave. Further, we obtain certain formulas for this type of solution in terms of the initial data. These solutions may in turn be useful in the study of more complex solutions, \textit{i.e.} nonlinear superposition of simple waves, and in the investigation of the global existence and uniqueness of those solutions as well as the nature of the gradient catastrophe. This topic is much better understood in the case where the Riemann wave problem reduces to the examination of a certain exterior differential system expressed in terms of Riemann invariants, known to be in involution in the sense of Cartan \cite{Cartan:1953}. In this paper, these theoretical considerations are systematically used to generate all nonlinear wave propagations admitted by the fluid dynamical system (\ref{eq:euler}) in (3+1) dimensions. A broad review of recent developments in this subject can be found in books such as A. Jeffrey \cite{Jeffrey:1976}, A. Madja \cite{Madja:1984}, Z.Peradsynski \cite{Peradzynski:1985}, B. Rozdestvenski and Y. Janenko\cite{Rozdestvenski:1983} and reference therein.
\paragraph{}In what follows we assume that all mappings, tensor fields and manifolds are smooth enough to make our considerations relevant and we use the summation convention unless otherwise stated.
\paragraph{}This paper is organized as follows. Section \ref{sec:2} contains a detailed account of rank-2 solutions expressed in terms of Riemann invariants. These results are used in Section \ref{sec:3} to formulate and solve the Riemann wave Cauchy problem for inhomogeneous systems. Section \ref{sec:4} contains a detailed account of the algebraic properties of the inhomogeneous Euler system (\ref{eq:euler}). In section \ref{sec:5}, we describe in detail a procedure for constructing solutions of the Euler equations (\ref{eq:euler}) and illustrate this procedure through examples.
\section{Rank-2 solutions}\label{sec:2}
In this section we consider the possibility of adapting the Riemann invariant approach for quasilinear systems of PDEs to the construction of a propagation of a simple wave allowed by an inhomogeneous system. This topic has already been discussed by the authors in \cite{Grundland:1974:a,Grundland:1974:b,GrundlandLamothe:2013,GrundlandLamothe:2014,GrundlandZelazny:1983}. However our present approach goes deeper into the algebraic and geometric aspects which will later enable us to obtain several new classes of exact solutions of the inhomogeneous system of fluid dynamics (\ref{eq:euler}). We will show that these solutions contain arbitrary functions of one variable.
\paragraph{}Recall that in order to construct simple waves of the inhomogeneous quasilinear system of partial differential equations (PDE)s in $n$ independent variables $x^\mu$ and $l$ unknown functions $u^i$
\begin{eqe}\label{eq:2.1}
\sum_{\mu=1}^n\sum_{j=1}^la^{s\mu}_j\pa{u^1,\ldots,u^l}\frac{\del u^j}{\del x^\mu}=b^s(u^1,\ldots,u^l),\quad s=1,\ldots, m,
\end{eqe}%
we must find two sets of $C^1$ functions $\lambda^1_1,\ldots,\lambda^1_n$, $\gamma_1^1,\ldots,\gamma_1^l:\mathcal{U}\rarrow \RR$ and $\lambda^0_1,\ldots,\lambda^0_n$, $\gamma_0^1,\ldots,\gamma_0^l:\mathcal{U}\rarrow \RR$ satisfying the system of algebraic equations
\begin{eqe}\label{eq:2.2}
a^{s\mu}_j(u^1,\ldots,u^l)\lambda^1_\mu\gamma^j _1=0,\qquad a^{s\mu}_j\pa{u^1,\ldots, u^l}\lambda^0_\mu\gamma^j_0=b^s(u^1,\ldots,u^l),
\end{eqe}%
where we denote by $\mathcal{U}$ and $X$ the spaces of dependent variables $u=\pa{u^1,\ldots,u^l}$ and independent variables $x=\pa{x^1,\ldots,x^n}$, respectively. To obtain these functions $\lambda=(\lambda_1,\ldots,\lambda_n)$, $\gamma=\pa{\gamma^1,\ldots,\gamma^l}$ and $\lambda^0=\pa{\lambda_1^0,\ldots,\lambda_n^0}$, $\gamma_0=\pa{\gamma_0^1,\ldots,\gamma_0^l}$ one proceeds by requiring that the $m\times l$ matrix $a^{s\mu}_j\lambda_\mu$ have rank less than $l$ and that the $m\times l$ matrix $a^{s\mu}_j\lambda^0 _\mu$ satisfy
\begin{eqe}\label{eq:2.3}
\rank\pa{a^{s\mu_j}\lambda}<l,\quad \rank\pa{a^{s\mu}_j\lambda^0_\mu,b^s}=\rank\pa{a^{s\mu}_j\lambda^0_\mu}
\end{eqe}%
at some generic point $u=(u^1,\ldots, u^l)\in \mathcal{U}$. These conditions obviously provide some algebraic restrictions on the functions $\lambda$ and $\lambda^0$. Suppose that we have obtained the two sets of functions $\lambda$ and $\lambda^0$ satisfying the above rank conditions for which we assume a linearly independent set of vectors at each point of some open subset $\mathcal{O}\subset \mathcal{U}$. For each $\lambda$ and $\lambda^0$ so obtained, we can solve the system (\ref{eq:2.2}) for $\gamma$ and $\gamma_0$, respectively.
\paragraph{Definition 2.1}
A rank-2 solution $u:X\rarrow \mathcal{U}$ of the inhomogeneous quasilinear system (\ref{eq:2.1}) is said to be a propagation of a simple wave on a simple state if the matrix ${\del u j}/{\del x^\mu}$ can be decomposed as
\begin{eqe}\label{eq:2.4}
\frac{\del u^j}{\del x^\mu}=\xi \gamma\otimes\lambda+\gamma_0\otimes \lambda^0,
\end{eqe}%
for each point $x\in X$, where $\otimes$ denotes the tensor product and $\xi$ is an arbitrary function of $x$.
\paragraph{}To show that the matrix ${\del u^j}/{\del x^\mu}$ has the correct decomposition for all indices $j$ and $\mu$, we multiply on the left by the matrix $a^{s\mu}_j(u)$ to obtain
\begin{eqe}\label{eq:2.5}
a^{s\mu}_j(u)\frac{\del u^j}{\del x^\mu}=\xi(x)a^{s\mu}_j\gamma_1^j\lambda^1_\mu+a^{s\mu}_j\gamma_0^j\lambda^0_\mu=b^s.
\end{eqe}%
Hence the system (\ref{eq:2.1}) holds because of the way that the so-called simple elements $\xi\gamma\otimes\lambda$ and $\gamma_0\otimes \lambda^0$ have been constructed. The physical meaning of the linear combination of these simple elements $\xi\gamma\otimes\lambda$ and $\gamma_0\otimes\lambda^0$ are quite different \cite{GrundlandZelazny:1983,Peradzynski:1971:planewave}. While the homogeneous element is usually attributed to certain waves, which can propagate in the medium, the inhomogeneous element leads to some special solution which will be called a simple state and which in general may not be attributed to a wave. In the literature \cite{Jeffrey:1976,Lighthill:1968,Madja:1984,Mises:1958,Peradzynski:1971:planewave,Peradzynski:1985,Riemann:1858,Rozdestvenski:1967,Rozdestvenski:1983,Whitham:1974,Zakharov:1998} solutions $u$ of $\rank\del u^j/\del x^\mu=1$ are actually called simples waves. We use the word "wave" for solutions which are interpreted as physical waves. For instance, elliptic inhomogeneous systems also have solutions of rank one, but we cannot call them "waves". Instead, they are called "modes" \cite{Peradzynski:1985}. This is why we have chosen to call solutions $u$ of the system (\ref{eq:2.1}) such that $du=\gamma_0\otimes \lambda^0$ "simple states". In this paper, we look for solutions of the form (\ref{eq:2.4}), where the matrix of the tangent mapping $\del u^i/\del x^\mu$ is the sum of a homogeneous and an inhomogeneous element. Note that a correct choice of the element of the form (\ref{eq:2.4}) leaves much freedom and requires us to study the structure of its components as well as the corresponding solution. A physical interpretation of this type of solution may be thought of as an interaction of waves with a medium in some state. The necessary and sufficient conditions for the existence of solutions of (\ref{eq:2.4}) were derived in \cite{Grundland:1974:a}. These conditions guarantee the propagation of a single wave solution described by the inhomogeneous quasilinear system (\ref{eq:2.1}). At this point let us summarize our approach for constructing rank-2 solutions of (\ref{eq:2.1}) subjected to (\ref{eq:2.4}). We make the assumption that we choose a holonomic system for the vector fields $\ac{\gamma_0,\gamma}$ by requiring a proper length for each vector such that
\begin{eqe}\label{eq:2.6}
\br{\gamma_0,\gamma}=0.
\end{eqe}%
This requirement means that there exists a parametrization of a surface $S$ immersed in the space of dependent variables $\mathcal{U}$
\begin{eqe}\label{eq:2.7}
u=f(r^0, r^1),
\end{eqe}%
which is obtained by solving the system of PDEs
\begin{eqe}\label{eq:2.8}
\frac{\del f^j}{\del r^i}=\gamma^j_i(f^1,\ldots,f^l),\quad i=0,1,
\end{eqe}%
where $r^0$ and $r^1$ are parameters along the respective integral trajectories of the vector fields $\gamma_0$ and $\gamma_1$ on $S$. Assuming that we have the parametric representation of the surface $S$ in the space $\mathcal{U}$, we consider the functions $f^\ast(\lambda^i_\mu)$, that is, the functions $\lambda^i_\mu(u)$ pulled back to the surface $S\subset\mathcal{U}$. The $\lambda^i_\mu(u)$ then become functions of the parameters $r^0$ and $r^1$ on $S$. In order to simplify the notation we denote $f^\ast(\lambda^j_\mu)$ by $\lambda^i_\mu(r^0,r^1)$. Thus, taking the differential $du(x)$ of the expression (\ref{eq:2.7}), we get
\begin{eqe}\label{eq:2.9}
du^j(x)=\frac{\del u^j}{\del x^\mu}dx^\mu=\frac{\del f^j}{\del r^i}\otimes dr^i.
\end{eqe}%
Comparing expressions (\ref{eq:2.4}) and (\ref{eq:2.9}) and assuming that the vectors $\gamma_0$ and $\gamma_1$ are linearly independent, we obtain a system of 1-forms
\begin{aleq}\label{eq:2.10}
dr^0(x)&=\lambda^0\pa{r^0(x),r^1(x)},\\
dr^1(x)&=\xi^1(x)\lambda^1\pa{r^0(x),r^1(x)},
\end{aleq}%
where $\lambda^0\wedge\lambda^1\neq0$ and $\xi^1(x)\neq0$.
This system of 1-forms is an involutive system in the sense of Cartan if the conditions \cite{Grundland:1974:b}
\begin{aleq}\label{eq:2.11}
\frac{\del \lambda^i}{\del r^j}&\in\operatorname{span}\ac{\lambda^0,\lambda^1},\quad i\neq j=1,2,\\
\frac{\del\lambda^0}{\del r^0}&\in\operatorname{span}\ac{\lambda^0}
\end{aleq}%
are satisfied. It was shown in \cite{Grundland:1974:a} that the conditions (\ref{eq:2.11}) are necessary and sufficient for the existence of a rank-2 solution of (\ref{eq:2.1}) describing a propagation of a simple wave on a simple state depending on two arbitrary analytic functions of one variable.
\paragraph{}In order to illustrate our method let us consider the case with two independent variables $t$ and $x$. After the elimination of the variable $\xi^1$ in the system (\ref{eq:2.10}), we obtain
\begin{aleq}\label{eq:r}
&i)\quad &&r^1_{,t}+v_1(r^0,r^1)r^1_{,x}=0,\\
&ii) &&r^0_{,t}=\lambda^0_1(r^0,r^1),\qquad r^0_{,x}=\lambda^0_2(r^0,r^1).
\end{aleq}%
Now we show that a solution of the system (\ref{eq:r}) describes the propagation of a single simple wave and we justify the notion "simple wave on a simple state" for this situation. It was proved \cite{Friedrich:1948,Mises:1958} that if the initial data is sufficiently small, then there exists a time interval $\br{t_0,T}$ in which the gradient catastrophe for the solution $r^1(t,x)$ of the system (\ref{eq:r}) does not occur since the function $r^1(t,x)$ is constant along the characteristic $C^{(1)}:dx/dt=v_1(r^1(t,x))$ of the system (\ref{eq:r}). If we choose in the space $X$ of independent variables the initial condition for the function $r^1$ in such a way that the derivative $r^1_{,x}$ has compact support
\begin{eqe}\label{eq:supp}
t=t_0: \operatorname{supp}r^1_{,x}(t_0,x)\subset \br{a,b},\quad a<b,
\end{eqe}%
then for arbitrary time $t_0<t<T$, $\operatorname{supp}r^1_{,x}(t,x)$ is contained in the strip between characteristics of the family $C^{(1)}$ passing through the ends of the interval $\br{a,b}.$ In this case the strip containing $\operatorname{supp}r^1(t,x)$ divides the remaining part of the space $X$ into two disjoint regions. In the region $G:=X\backslash\ac{\operatorname{supp}r^1(t,x)}$ the solution $r^0(t,x)$ of the system (\ref{eq:r}) is described by the simple state. In this region $r^1_{,x}=0$ holds and the solution $r^0(t,x)$ satisfies equation (\ref{eq:r}.ii) with $r^1=r^1_0=\text{const}$. From the compatibility of the equations (\ref{eq:r}.ii), we obtain
\begin{eqe}\label{eq:colin}
\lambda^0_{,r^0}\wedge \lambda^0=0
\end{eqe}%
which means that the direction of $\lambda^0$ does not depend on the variable $r^0$, so it is constant on $G$. Choosing the parametrization of the curve $u=f(r^0,r^1_0)$, in such a way that the covector $\lambda^0$ does not depend on the parameter $r^0$, we can express the solution on region $G$ in the form of a simple state, \textit{i.e.}
\begin{eqe}\label{eq:SS}
u^j=f^j(r^0,r^1),\quad\text{where}\ \frac{df^j}{dr^0}=\gamma^j(r^0,r^1_0),\quad j=1,\ldots,l,\\
r^0=\lambda_0^0 t+\lambda^0_1 x,
\end{eqe}%
where $\lambda^0=\pa{\lambda^0_0,\lambda^0 _1}$ is a constant vector.
\paragraph{}For the general case (an arbitrary number of independent variables), it is convenient from the computational point of view to express the integral conditions (\ref{eq:2.11}) in explicit form because there exist functions $\alpha_i$ and $\beta_i:S\rarrow\RR$ such that
\begin{aleq}\label{eq:2.12}
&(a)\quad &&\frac{\del \lambda^0}{\del r^0}=\alpha_0 (r^0,r^1)\lambda^0,\qquad (b)\quad\frac{\del \lambda^0}{\del r^1}=\alpha_1(r^0,r^1)\lambda^1\\\
&(c)\quad &&\frac{\del \lambda^1}{\del r^0}=\beta_0 (r^0,r^1)\lambda^0+\beta_1(r^0,r^1)\lambda^1
\end{aleq}%
hold. In order to construct rank-2 simple wave solutions of the inhomogeneous system (\ref{eq:2.1}) we consider two separate cases, namely, when the coefficient $\alpha_1$ of $r^0$ and $r^1$ in (\ref{eq:2.12}) does not vanish anywhere and when this coefficient $\alpha_1$ is identically equal to zero.
\subsection{The case when $\alpha_1\neq 0$}In this case, equation (\ref{eq:2.12}.c) is a consequence of equations (\ref{eq:2.12}.a) and (\ref{eq:2.12}.b):
$$\frac{\del \lambda^1}{\del r^0}=\frac{\del}{\del r^0}\pa{\frac{1}{\alpha_1}\frac{\del \lambda^0}{\del r^1}}=\frac{1}{\alpha_1}\frac{\del \alpha_0}{\del r^1}\lambda^0+\pa{\alpha_0-\frac{1}{\alpha_1}\frac{\del \alpha_1}{\del r^0}}\lambda^1.$$%
Let the function $\varphi$ be the solution of the equation $\del \varphi/\del r^0=\alpha_0(r^0,r^1)$. Then we obtain
\begin{aleq}\label{eq:2.13}
&\lambda^0(r)=\lambda^1(r^1)\exp\varphi(r),\\
&\lambda^1(r)\sim\frac{\del \lambda^0}{\del r^1}\sim \pa{\frac{\del \varphi}{\del r^1}\lambda^1(r^1)+\dot{\lambda^1}(r^1)},
\end{aleq}%
where we have denoted $r=(r^0,r^1)$ and $\dot{\lambda}=d\lambda(r^s)/dr^s$. So we can study the system (\ref{eq:2.10}) in the form
\begin{aleq}\label{eq:2.14}
dr^0&=\lambda(r^1)\exp(\varphi(r)),\quad \text{where}\ \lambda(r^1)\wedge\dot{\lambda}(r^1)\neq 0,\\
dr^1&=\xi\pa{\frac{\del \varphi}{\del r^1}\lambda(r^1)+\dot{\lambda}(r^1)},
\end{aleq}%
for which the integrability conditions (\ref{eq:2.12}) are automatically satisfied. According to the definition presented in \cite{Jeffrey:1964}, the result of a propagation of a simple wave on a simple state is strictly nonlinear if for all $i,j\in\ac{0,1}$ such that $i\neq j$ the conditions $\del \lambda^i/\del r^j\neq 0$ hold. This situation takes place when we have $\alpha_1\neq 0$ in the expression (\ref{eq:2.11}) and consequently the wave vectors $\lambda^0$ and $\lambda^1$ take the forms (\ref{eq:2.13}).
\paragraph{}We now show that solutions of the system (\ref{eq:2.14}) can be expressed in the implicit form
\begin{aleq}\label{eq:2.15}
\lambda_{\mu}(r^1)x^\mu&=\Psi^0(r^0,r^1),\\
\dot{\lambda}_{\mu}(r^1)x^\mu&=\Psi^1(r^0,r^1).
\end{aleq}%
To simplify the formulae, we use the following notation
$$v^0=\lambda_\mu(r^1)x^\mu,\quad v^1=\dot{\lambda}_\mu(r^1)x^\mu,\quad v^2=\ddot{\lambda}_\mu(r^1)x^\mu.$$%
Differentiating equations (\ref{eq:2.15}), we obtain
$$\begin{aligned}
\dot{\lambda}_\mu x^\mu dr^1+\lambda&=\frac{\del \Psi^0}{\del r^0}dr^0+\frac{\del \Psi^0}{\del r^1}dr^1,\\
\ddot{\lambda}_\mu x^\mu dr^1+\dot{\lambda}&=\frac{\del \Psi^1}{\del r^0}dr^0+\frac{\del \Psi^1}{\del r^1}dr^1,
\end{aligned}$$%
or equivalently, written in a matrix form,
\begin{aleq*}
\pa{\begin{array}{cc}
      \frac{\del \Psi^0}{\del r^0}, & \frac{\del \Psi^0}{\del r^1}-v^1 \\
      \frac{\del\Psi^1}{\del r^0}, & \frac{\del \Psi^1}{\del r^1}-v^2
    \end{array}
}\pa{\begin{array}{c}
     dr^0 \\
     dr^1
    \end{array}}=\pa{\begin{array}{c}
     \lambda \\
     \dot{\lambda}
    \end{array}}.\end{aleq*}%
So we have
\begin{eqe}\label{eq:2.18}
\pa{\begin{array}{c}
     dr^0 \\
     dr^1
    \end{array}}=\frac{1}{W}\pa{\begin{array}{cc}
                       \frac{\del \Psi^1}{\del r^1}-v^2, & -\frac{\del \Psi^0}{\del r^1}+v^1 \\
                       -\frac{\del \Psi^1}{\del^0}, & \frac{\del\Psi^0}{\del r^0}
                     \end{array}
    }\pa{\begin{array}{c}
     \lambda \\
     \dot{\lambda}
    \end{array}},
\end{eqe}%
where
$$W=\pa{\frac{\del \Psi^1}{\del r^1}-v^2}\frac{\del \Psi^0}{\del r^0}+\pa{-\frac{\del\Psi^0}{\del r^1}+v^1}\frac{\del \Psi^1}{\del r^0}\neq 0.$$%
Inserting (\ref{eq:2.18}) into the system of equations (\ref{eq:2.14}) we get
\begin{aleq}\label{eq:2.19}
&(a)\quad &&\frac{\del \Psi^0}{\del r^1}=\Psi^1,\\
&(b) &&\frac{\del \Psi^1}{\del r^0}+\frac{\del \varphi}{\del r^1}\frac{\del \Psi^0}{\del r^0}=0,
\end{aleq}%
and
$$
\frac{\frac{\del \Psi^1}{\del r^1}-v^2}{\pa{\frac{\del \Psi^1}{\del r^1}-v^2}\frac{\del \Psi^0}{\del r^0}-\pa{\frac{\del \Psi^0}{\del r^1}-\Psi^1}\frac{\del \Psi^1}{\del r^0}v^2}\equiv \exp\varphi.
$$
So we have
$$\begin{aligned}
\frac{\del \Psi^0}{\del r^0}&=\exp-\varphi,\\
\frac{\del \Psi^1}{\del r^1}&=\pa{\frac{\del \Psi^0}{\del r^0}\frac{\del \Psi^1}{\del r^1}-\pa{\frac{\del \Psi^0}{\del r^1}-\Psi^1}\frac{\del \Psi^1}{\del r^0}}\exp\varphi.
\end{aligned}$$%
From these equations we find
$$\pa{\frac{\del \Psi^0}{\del r^1}-\Psi^1}\frac{\del \Psi^1}{\del r^0}=0.$$%
This is a consequence of equation (\ref{eq:2.19}.a). Finally, we obtain the system of differential equations for the unknown functions $\psi^0$ and $\psi^1$
\begin{aleq}\label{eq:2.21}
&(a) \quad&&\frac{\del \Psi^0}{\del r^1}=\Psi^1,\\
&(b) &&\frac{\del \Psi^0}{\del r^0}=\exp(-\varphi),\\
&(c) &&\frac{\del \Psi^1}{\del r^0}+\frac{\del \varphi}{\del r^1}\exp(-\varphi)=0.
\end{aleq}%
Note that equation (\ref{eq:2.21}.c) is a closure condition for the one-form $\omega=\exp(-\varphi)dr^0+\Psi^1dr^1$, that is the integrability condition of the system (\ref{eq:2.12}.a) and (\ref{eq:2.12}.b) for the function $\Psi^0$. The general solution of equation (\ref{eq:2.12}.c) has the form
$$\Psi^1(r)=-\int_0^{r^0}\frac{\del \varphi(\xi,r^1)}{\del r^1}\exp(-\varphi(\xi,r^1))d\xi+\dot{\Phi}(r^1),$$%
where the constant of integration with respect to $r^0$, depending on $r^1$, is, for convenience, denoted by $\dot{\Phi}(r^1)$. Given the function $\Psi^1(r)$ we can solve the system of equations (\ref{eq:2.21}.a) and (\ref{eq:2.21}.b). Integrating the closed form $\omega$, we get
$$\Psi^0(r)=\int_{\gamma_0,R}\omega=\int^{r^1}_0\Phi(r^1)dr^1+\int_0^{r^0}\exp(-\varphi(\xi,r^1))d\xi,$$%
where $\gamma_{0,R}$ is a broken line with vertices 0, $(0,r^1)$, $r$. Thus the general solution of the system (\ref{eq:2.21}) has the form
\begin{aleq}\label{eq:2.22}
\Psi^0(r)&=\Phi(r^1)+\int_0^{r^0}\exp(-\varphi(\xi,r^1))d\xi,\\
\Psi^1(r)&=\dot{\Phi}(r^1)-\int_0^{r^0}\frac{\del \varphi(\xi,r^1)}{\del r^1}\exp(-\varphi(\xi,r^1))d\xi.
\end{aleq}%
So we have the following proposition:
\paragraph{Proposition 2.1} All solutions of the system (\ref{eq:2.14}) can be obtained by solving the implicit system of equations
\begin{aleq}\label{eq:2.23}
\lambda_\mu(r^1)x^\mu&=\Phi(r^1)+\int_0^{r^0}\exp(-\varphi(\xi,r^1))d\xi,\\
\dot{\lambda}_\mu(r^1)x^\mu&=\dot{\Phi}(r^1)-\int_0^{r^0}\frac{\del \varphi(\xi,r^1)}{\del r^1}\exp(-\varphi(\xi,r^1))d\xi,
\end{aleq}%
with respect to the variables $r^0$, $r^1$. Here $\Phi(\cdot)$ is an arbitrary differentiable function of $r^1$.
\paragraph{Proof.} In view of our computations, it is sufficient to show that any nondegenerate (that is with $\xi\not\equiv 0$) solution of the system (\ref{eq:2.14}) can be expressed in the form (\ref{eq:2.23}). Indeed, if (\ref{eq:2.23}) is satisfied, then
$$\begin{aligned}
d\pa{\lambda_\mu(r^1)x^\mu}&=\dot{\lambda}_\mu(r^1)x^\mu dr^1+\lambda(r^1)\\
&=\exp(-\varphi)dr^0+\pa{\dot{\lambda}_\mu x^\mu}dr^1,\\
d\pa{\dot{\lambda}_\mu(r^1)x^\mu}&=\ddot{\lambda}(r^1)x^\mu dr^1+\dot{\lambda}(r^1)\\
&=\frac{\del \varphi}{\del r^1}\exp(-\varphi)dr^0+\pa{\ddot{\lambda}_\mu x^\mu+\frac{1}{\xi^1}}dr^1.
\end{aligned}$$%
So both functions $\lambda_\mu(r^1)x^\mu$ and $\dot{\lambda}(r^1)_\mu x^\mu$ can be expressed as functions of $r^0(x)$ and $r^1(x)$.\\
\makebox[\textwidth][r]{$\Box$}
\subsection{The case when $\alpha_1=0$}In this case it follows from equations (\ref{eq:2.12}.a) and (\ref{eq:2.12}.b) that
\begin{eqe}\label{eq:2.star1}
\lambda^0(r)=C\exp\varphi(r^0),\quad \text{where}\ C\in X^{\ast},
\end{eqe}%
where $X^\ast$ is the space of linear forms and $\varphi(r^0)$ is a differentiable function of one variable. By virtue of equation (\ref{eq:2.12}.c) for arbitrary functions $\zeta=\zeta(r)$ and $\chi=\chi(r)$ we have
$$\frac{\del}{\del r^0}\pa{\lambda^1\exp(-\zeta-\chi)C}=\pa{\beta_0\exp\pa{-\zeta+\varphi}-\frac{\del \chi}{\del r^0}}C+\pa{\beta_1-\frac{\del \zeta}{\del r^0}}\exp\pa{-\zeta}\lambda^1.$$%
So for the quantities $\zeta$, $\chi$, we take the solutions of the system
$$\frac{\del \zeta}{\del r^0}=\beta_1,\quad \frac{\del \chi}{\del r^0}=\beta^0\exp(-\zeta+\varphi),$$%
(which always exist locally), and obtain
\begin{eqe}\label{eq:2.star2}
\lambda^1(r)\exp(-\zeta(r))=\chi(r)C+A(r^1),
\end{eqe}%
where $A$ is a differentiable function of $r^1$.
\paragraph{}In the case when $\alpha_1=0$ the wave vectors $\lambda^0$ and $\lambda^1$ are given by the expressions (\ref{eq:2.star1}) and (\ref{eq:2.star2}). Hence, the superposition is nonlinear since for some $i,j\in\ac{0,1}$ such that $i\neq j$ the wave vectors $\lambda^0$ and $\lambda^1$ satisfy the conditions $\del \lambda^i/\del r^j\neq 0$. Thus, in this case, system (\ref{eq:2.10}) has the form
\begin{aleq}\label{eq:2.24}
&(a)\quad &&d\Psi^0(r^0)=C,\\
&(b) &&dr^1=\xi^1\pa{\chi(r)C+A(r^1)},
\end{aleq}%
where $\Psi^0(r^0):=\int_0^{r^0}\exp(-\varphi(\xi))d\xi$. From equation (\ref{eq:2.24}.a) it follows immediately that $\Psi^0(r^0)=c_\mu x^\mu+a_0$, $a_0\in\RR$. We notice that if the variable $r^1$ satisfies equation (\ref{eq:2.22}.b) in which $r^0=({\Psi^0})^{-1}(c_\mu x^\mu+a_0)$, then
$$\begin{gathered}
d\pa{\int_0^{r^0}\chi(r^0,r^1)dr^0+A_\mu(r^1)x^\mu}=\pa{\int_0^{r^0}\frac{\del \chi(r^0,r^1)}{\del r^1}dr^0+\dot{A}_\mu(r^1)x^\mu}dr^1\\
+A(r^1)+\chi(r^0,r^1)C\sim dr^1.
\end{gathered}$$%
So we have
$$\int_0^{r^0}\chi(\xi,r^1)\exp(-\varphi(\xi))d\xi+A_\mu(r^1)x^\mu=\psi^1(r^1),$$%
where $\psi^1$ is a differentiable function of $r^1$. We have the following proposition:
\paragraph{Proposition 2.2}The general integral of the inhomogeneous system (\ref{eq:2.24}) has the implicit form
\begin{eqe}\label{eq:2.25}
c_{\mu}x^\mu+a_0=\Psi^0(r^0),\quad \int_0^{r^0}\chi(\xi,r^1)\exp(-\varphi(\xi))d\xi+A_\mu(r^1)x^\mu=\psi^1(r^1),
\end{eqe}%
where $\psi^1(\cdot)$ is an arbitrary differentiable function of $r^1$, and
$$\Psi^0(r^0)=\int_0^{r^0}\exp(-\varphi(\xi))d\xi.$$%
\paragraph{Remark 2.1} The above solutions can be generalized to the case of many simple waves in the inhomogeneous system
\begin{aleq}\label{eq:2.26}
dr^0&=C\exp\varphi(r^0),\\
dr^s&=\xi^s\pa{\chi^s(r^0,r^s)C+A^s(r^s)},\ s=1,\ldots,p,
\end{aleq}%
where $C$, $A^1(r^1),\ldots,A^p(r^p)$ are linearly independent 1-forms. The general integral of the system (\ref{eq:2.26}) has the form
\begin{gaeq*}
C_{\mu}x^\mu+a_0=\int_0^{r^0}\exp(-\varphi(r^0))dr^0,\\
\int_0^{r^0}\chi^s(\xi,r^s)\exp(-\varphi(\xi))d\xi+A_\mu^s(r^s)x^\mu=\psi^s(r^s),
\end{gaeq*}%
where $\Psi^1(\cdot),\ldots,\Psi^s(\cdot)$ are arbitrary functions of their arguments.
\section{Formulation of the Cauchy problem for the case of the propagation of a simple wave  on a simple state}\label{sec:3}
Let us now study an example of the formulation of the Cauchy problem for the Pfaffian system of the form
\begin{aleq}\label{eq:II:1}
dr^0&=\xi^0\lambda^0(r),\ \text{where}\ &&\lambda^0(r)=\lambda(r^1)\exp\varphi(r),\\
dr^1&=\xi^1\lambda^1(r), &&\lambda^1(r)=\lambda(r^1)\frac{\del \varphi(r)}{\del r^1}+\dot{\lambda}(r^1).
\end{aleq}%
According to the preceding section, the integrability conditions (\ref{eq:2.12}) are automatically satisfied. We are looking for a solution in the form
\begin{eqe}\label{eq:II:2}
\lambda_\mu(r^1)x^\mu=G^0(r),\qquad \dot{\lambda}_\mu(r^1)x^\mu=G^1(r).
\end{eqe}%
Then we have
$$\begin{aligned}
\lambda+\dot{\lambda}_\mu x^\mu dr^1&=\frac{\del G^0}{\del r^0}dr^0+\frac{\del G^0}{\del r^1}dr^1,\\
\dot{\lambda}+\ddot{\lambda}_\mu x^\mu dr^1&=\frac{\del G^1}{\del r^0}dr^0+\frac{\del G^1}{\del r^1}dr^1,
\end{aligned}$$%
from which we get
$$\begin{aligned}
dr^0\sim \pa{\frac{\del G^1}{\del r^1}-\ddot{\lambda}_\mu x^\mu}\lambda+\pa{-\frac{\del G^0}{\del r^1}+ \dot{\lambda}_\mu x^\mu}\dot{\lambda},\\
dr^1\sim\pa{-\frac{\del G^1}{\del r^0}\lambda+\frac{\del G^0}{\del r^0}\dot{\lambda}}.
\end{aligned}$$%
So, we have
$$\frac{\del G^0}{\del r^1}=G^1,\quad \det\left|\begin{array}{cc}
                                                   -\frac{\del G^1}{\del r^0}, & \frac{\del G^0}{\del r^0} \\
                                                   \frac{\del \varphi}{\del r^1}, & 1
                                                 \end{array}
\right|=0,$$%
and we obtain the system of equations
\begin{eqe}\label{eq:II:3}
\frac{\del G^0}{\del r^1}=G^1,\quad \frac{\del G^1}{\del r^0}+\frac{\del \varphi}{\del r^1}\frac{\del G^0}{\del r^0}=0.
\end{eqe}%
It follows that $\frac{\del^2G^0}{\del r^1\del r^0}+\frac{\del \varphi}{\del r^1}\frac{\del G^0}{\del r^0}=0$ so $\frac{\del }{\del r^1}\pa{\frac{\del G^0 }{\del r^0}\exp\varphi}=0$. Then $\frac{\del G^0(r)}{\del r^0}=a(r^0)\exp(-\varphi(r))$, where $a(r^0)$ is an arbitrary differentiable function of $r^1$. Finally we have
\begin{aleq}\label{eq:II:4}
G^0(r)&=\Phi(r^1)+\int_0^{r^0}a(\xi)\exp(-\varphi(\xi,r^1))d\xi,\\
G^1(r)&=\dot{\Phi}(r^1)-\int_0^{r^0}a(\xi)\frac{\del \varphi(\xi,r^1)}{\del r^1}\exp(-\varphi(\xi,r^1))d\xi.
\end{aleq}%
Suppose that, on some curve $\Gamma\in X$, the values of the functions are given by $r|_\Gamma=\pa{r^0|_\Gamma,r^1|_\Gamma}$. Let us assume that the functions $r^0|_\Gamma$ and $r^1|_\Gamma$ are invertible. Then, as a parameter of this curve, we can choose the value $r^1$, \textit{i.e.}
$$\Gamma=\ac{x=\eta^1(r^1)},\quad r\circ\eta^1(r^1)=\pa{\rho^0(r^1),r^1},$$%
or the value $r^0$, \textit{i.e. }
$$\Gamma=\ac{x=\eta^0(r^0)},\quad r\circ\eta^0(r^0)=\pa{r^0,\rho^1(r^0)}.$$%
Then the functions $r^0\mapsto \rho^1(r^0)$ and $r^1\mapsto \rho^0(r^1)$ are the inverses of each other. Moreover, by virtue of the identity $\eta^1(r^1)=\eta^0(\rho^0(r^1))$ we have
\begin{eqe}\label{eq:II:5}
\dot{\eta}^1(r^1)=\dot{\eta}^0(r^0)\dot{\rho}^0(r^1).
\end{eqe}%
Inserting this into equations (\ref{eq:II:2}) and (\ref{eq:II:4}), we get
\begin{aleq}\label{eq:II:6}
&(a) \quad &&\lambda_\mu(r^1)\eta^{1\mu}(r^1)&&=\Phi(r^1)+\int_0^{\rho^0(r^1)}a(\xi)\exp(-\phi(\xi,r^1))d\xi,\\
&(b) &&\dot{\lambda}_\mu(r^1)\eta^{1\mu}(r^1)&&=\dot{\Phi}(r^1)-\int_0^{\rho^0(r^1)}a(\xi)\frac{\del \varphi(\xi,r^1)}{\del r^1}\exp(-\varphi(\xi,r^1))d\xi.
\end{aleq}%
Differentiating equation (\ref{eq:II:6}.a) and subtracting equation (\ref{eq:II:6}.b), we get
$$
\lambda_\mu(r^1)\dot{\eta}^{1\mu}(r^1)=\dot{\rho}^0(r^1)a(\rho^0(r^1))\exp(-\varphi(\rho^0(r^1),r^1)).
$$
Taking equation (\ref{eq:II:5}) into account, it follows that
\begin{eqe}\label{eq:II:8}
a(r^0)=\lambda_\mu(\rho^1(r^0))\dot{\eta}^{0\mu}(r^0)\exp\varphi(r^0,\rho^1(r^0)),
\end{eqe}%
and after inserting (\ref{eq:II:8}) into equation (\ref{eq:II:6}.a), we have
\begin{eqe}\label{eq:II:9}
\Phi(r^1)=\lambda_\mu(r^1)\eta^{1\mu}(r^1)-\int_0^{\rho^0(r^1)}\lambda_\mu(\rho^1(\xi))\dot{\eta}^{0\mu}(\xi)\exp\pa{\varphi(r^1,\rho^1(r^0))-\varphi(\xi,r^1)}d\xi.
\end{eqe}%
Thus we get the following proposition:
\paragraph{Proposition 3.1} If on some curve $\Gamma\in X$ the Cauchy conditions $r^0|_\Gamma$ and $r^1|_\Gamma$ are given for the equation (\ref{eq:II:1}) in such a way that:
\begin{itemize}
\item[1)] the functions $r^0|_\Gamma$ and $r^1|_\Gamma$ are strongly monotonic,
\item[2)] for $x\in \Gamma$ a vector tangent to the curve $\Gamma$ does not belong to the annihilator of the forms $\lambda^0(r(x))$ and $\lambda^1(r(x))$,
 \end{itemize}%
 then there exists a tubular neighborhood of the curve $\Gamma$ for which the Cauchy problem of (\ref{eq:II:1}) has (locally) exactly one rank-2 solution. This solution represents a simple wave on a simple state.
 \paragraph{Proof.}It is enough to show that, for the functions $G^0$ and $G^1$ defined by formulae (\ref{eq:II:4}), (\ref{eq:II:8}) and (\ref{eq:II:9}), the system (\ref{eq:II:2}) can be solved in the neighborhood of an arbitrary point $x$ on the curve $\Gamma$. The conditions of local solvability have the form
 $$\begin{gathered}
 0\neq \operatorname{det}\left|\begin{array}{cc}
                                 \frac{\del G^0}{\del r^0}, & \frac{\del G^0}{\del r^1}-\dot{\lambda}_\mu x^\mu \\
                                 \frac{\del G^1}{\del r^0}, & \frac{\del G^1}{\del r^1}-\ddot{\lambda}_\mu x^\mu
                               \end{array}
 \right|=\left|\begin{array}{cc}
                 \frac{\del G^0}{\del r^0}, & 0 \\
                 \frac{\del G^1}{\del r^0}, & \frac{\del G^1}{\del r^1}-\ddot{\lambda}_\mu x^\mu
               \end{array}
 \right|=\frac{\del G^0}{\del r^0}\pa{\frac{\del G^1}{\del r^1}-\ddot{\lambda}_\mu x^\mu},
 \end{gathered}$$%
 where $x=\eta^0(r^0)\in \Gamma$. We have to prove, that $\frac{\del G^0}{\del r^1}\neq 0$ and $\frac{\del G^1}{\del r^1}-\ddot{\lambda}_\mu x^\mu\neq 0$. By virtue of (\ref{eq:II:4}.a) and (\ref{eq:II:8}), we have
 \begin{aleq}\label{eq:II:10}
 \frac{\del G^0}{\del r^0}&=a(r^1)\exp\pa{-\varphi(r^0,r^1)}=\lambda_\mu(r^1)\dot{\eta}^{0\mu}(r^0)\exp\pa{-\varphi(r^0,r^1)+\varphi(r^0,r^1)}\\
 &=\lambda_\mu(r^1)\dot{\eta}^{0\mu}\neq 0,
 \end{aleq}%
 because, by the assumption (\ref{eq:II:10}), the tangent vector $\dot{\eta}^0(r^0)$ does not belong to the annihilator of $\lambda(r^1)$. Similarly, for equations (\ref{eq:II:6}.b), (\ref{eq:II:3}.b) and (\ref{eq:II:10}), we have
 $$\begin{aligned}
 \frac{\del G^1}{\del r^1}-\ddot{\lambda}_\mu x^\mu&=\frac{d}{dr^1}\pa{G^1(\rho^0(r^1),r^1)}-\frac{\del G^1}{\del r^0}\dot{\rho}^0(r^0)-\ddot{\lambda}_\mu x^\mu\\
 &=\dot{\lambda}_\mu\dot{\eta}^{1\mu}+\frac{\del \varphi}{\del r^1}\frac{\del G^0}{\del r^0}=\pa{\frac{\del \varphi}{\del r^1}\lambda_\mu+\dot{\lambda}_\mu}\dot{\eta}^{1\mu}\neq 0,
 \end{aligned}$$%
 since, by the assumption (2), the tangent vector $\dot{\eta}^1$ does not vanish on $\lambda^1=(\del \varphi/\del r^1)\lambda +\dot{\lambda}$.\\
\makebox[\textwidth][r]{$\Box$}
\subsection{The case when $\alpha_1\neq 0$}In the case of an inhomogeneous system according to (\ref{eq:2.7}), we have
$$
dr^0=\lambda(r^1)\exp\varphi(r),\qquad dr^1=\xi^1\pa{\frac{\del \varphi(r)}{\del r^1}\lambda(r^1)+\dot{\lambda}(r^1)},
$$
and the solution is also determined by formulae (\ref{eq:II:2}) and (\ref{eq:II:4}) with the additional restriction $a(r^0)\equiv1$ on the arbitrary functions. From equation (\ref{eq:II:8}), we see that in this case, along the curve $\Gamma$, we can give only one function, for example $r^0|_\Gamma$ and then the other may be computed from the restriction $a(r^0)\equiv 1$. Thus, we have
\paragraph{Proposition 3.2:}Let the function $r^0|_\Gamma$ be given along some curve $\Gamma$. Let us assume that:
\begin{itemize}
\item[(1)] the function $r^0|_\Gamma$ is monotonic;
\item[(2)] the equation $\lambda^0(r^0,r^1)\circ \dot{\eta}^0(r^0)=1$ (where $x=\eta^0(r^0)$) is the equation of the curve $\Gamma$ parameterized by $r^0|_\Gamma$) which allows us to determine uniquely the value $r^1=r^1|_\Gamma$ along the curve $\Gamma$;
\item[(3)] the values $r^0|_\Gamma$ and $r^1|_\Gamma$ determined in this way satisfy the transversality condition with respect to the form $\lambda^1(r|_\Gamma)$;
\end{itemize}
then there exists a tubular neighborhood of the curve $\Gamma$ for which the inhomogeneous Cauchy problem (\ref{eq:II:1}) has (locally) exactly one rank-2 solution. This solution $u=f(r^0,r^1)$ represents a simple wave on a simple state.
\subsection{The case when $\alpha_1=0$}
Suppose that we are given the curve $\Gamma=\ac{x=h(t)}$ and that the value $r^1|_\Gamma$ is determined by the function $\rho^1(t)$. Let us assume that the function $r^1|_\Gamma$ is monotonic, so that we can choose it to be the parameter on the curve $r^1(h(r^1))=r^1$. Additionally, suppose that we are given the quantity $r^0(h(r_0^1))=\rho^0_0$. Then it follows from equation (\ref{eq:2.25}.a) that
\begin{eqe}\label{eq:II:12}
r^0=r^0(h(r^1))=(\Psi^0)^{-1}\pa{C h(r^1)+\rho^0_0-C h(r_0^1)}.
\end{eqe}%
From equation (\ref{eq:2.25}.b), we can determine the value of the function $\Psi$ of $r^1$, \textit{i.e.}
\begin{eqe}\label{eq:II:13}
\Psi^1(r^1)=\int_0^{C(h(r^1))-h(r_0^1)+\rho^0_0}\chi((\Psi^0)^{-1}(\xi),r^1)d\xi+A(r^1)h(r_1).
\end{eqe}%
\paragraph{}We look for the condition of local solvability of the system (\ref{eq:2.25}). Because we can always find the value of $r^0=r^0(x)$ via equation (\ref{eq:2.25}), this condition has the form
\begin{aleq}\label{eq:II:14}
0\neq&\frac{\del }{\del r^1}\int_0^{Cx+C_0}\chi\pa{(\Psi^0)^{-1}(\xi),r^1}d\xi+A(r^1)x-\Psi^1(r^1)\\
&=\int_0^{Cx+C_0}\frac{\del \chi\pa{(\Psi^0)^{-1}(\xi,r^1)}}{\del r^1}d\xi+\dot{A}(r^1)x-\dot{\Psi}^1(r^1).
\end{aleq}%
Differentiating (\ref{eq:II:13}), we have
$$\dot{\Psi}^1(r^1)=C \dot{h}(r^1)\chi(r^0,r^1)+\int_0^{C x-C_0}\frac{\del \chi((\Psi^0)^{-1}(\xi),r^1)}{\del r^1}d\xi +\dot{A}(r^1)h(r^1)+A(r^1)\dot{h}(r^1).$$%
Thus the condition (\ref{eq:II:14}) gives us
$$0\neq Ch(r^1)\chi(r^0,r^1)+A(r^1)\dot{h}(r^1)=\lambda^1(r^0,r^1)\dot{h}(r^1).$$%
Thus we have shown:
\paragraph{Proposition 3.3} If along the curve $\Gamma\subset E$, we are given the value of $r^1|_\Gamma$, which is assumed to be a monotonic function, and if for the value $r^0|_\Gamma$ determined from the formula (\ref{eq:II:12}), the tangent vector to the curve $\Gamma$ does not belong to the annihilator of the form $\lambda^1(r)|_\Gamma$, then the inhomogeneous Cauchy problem has exactly one solution in some neighborhood of the curve $\Gamma$.
\section{Algebraic properties of the inhomogeneous Euler equations}\label{sec:4}
The algebraic aspect of the inhomogeneous Euler system (\ref{eq:euler}) has already been presented in \cite{Grundland:1974:b}. However, our approach goes deeper into a systematic use of the Riemann invariants structure in order to generate several classes of solutions of the system (\ref{eq:euler}).
\paragraph{}Here we treat the space of independent variables $X\subset \RR^4$ as the classical space-time, where each of its points has coordinates $(t,\vec{x})$ and the space of unknown functions $U\subset\RR^5$ has the coordinates $(\rho, p ,\vec{v})$. Let us denote by $\lambda=\pa{\lambda_0, \vec{\lambda}}$, where $\vec{\lambda}\in \RR^3$, the vector field $\lambda=\lambda_\nu dx^\nu$ which belongs to the space of linear forms $X^\ast$ and by $\gamma_1=\pa{\gamma_\rho,\gamma_p,\vec{\gamma}}$ (where $\vec{\gamma}=\pa{\gamma^1,\gamma^2,\gamma^3}$) an element of the tangent space $T_u\mathcal{U}$, where $\rho\rarrow\gamma_\rho$, $p\rarrow\gamma_p$ and $\vec{v}\rarrow\vec{\gamma}$. Algebraic equations which determine simple homogeneous and inhomogeneous elements for the equations (\ref{eq:euler}) are of the form
\begin{aleq}\label{eq:4.1}
&\rho\delta|\vec{\lambda}|\vec{\gamma}+\gamma_p\vec{\lambda}=\rho\pa{\vec{g}-\vec{\Omega}\times\vec{v}},\\
&\delta|\vec{\lambda}|\gamma_\rho+\rho\vec{\gamma}\cdot\vec{\lambda}=0,\\
&\delta|\vec{\lambda}|\pa{\rho\gamma_p-\kappa p\gamma_\rho}=0,
\end{aleq}%
where we use the following notation for the derivative in terms of simple elements
\begin{eqe}\label{eq:4.2}
\frac{\del}{\del t}+\vec{v}\cdot\nabla\rarrow \delta|\vec{\lambda}|=\lambda_0+\vec{v}\cdot{\vec{\lambda}}.
\end{eqe}%
The function $\delta|\vec{\lambda}|$ has a physical interpretation. It describes the velocity of propagation of a disturbance relative to the fluid. Due to equation (\ref{eq:2.3}), there exists a nontrivial solution $\gamma_0\not\equiv 0$ if and only if the following conditions hold:
\begin{aleq}\label{eq:4.3}
&1 \quad E^0:\ \delta_{E^0}=\delta|\vec{\lambda}|=0\text{ and }\gamma_p=0,\\
&2 \quad A^0_{\epsilon}:\ \delta_{A_\epsilon}=\delta|\vec{\lambda}|=\epsilon\pa{\frac{\kappa p}{\rho}}^{\frac{1}{2}}\text{ and }\vec{\lambda}\cdot\pa{\vec{g}-\vec{\Omega}\times\vec{v}}=0,\quad \epsilon=\pm 1,\\
&3\quad H^0:\ \delta_{H^0}=\delta|\vec{\lambda}|\neq\left\{\begin{aligned}&0\\ &\epsilon\pa{\frac{\kappa p}{\rho}}^{\frac{1}{2}},\quad \epsilon=\pm 1.\end{aligned}\right.
\end{aleq}%
In what follows we use the notation $\epsilon=\pm 1$. The equations (\ref{eq:4.3}.1) and (\ref{eq:4.1}) determine the inhomogeneous entropic element which is denoted by $E^0$. This condition leads to the following inhomogeneous entropic element
$$
E^0:\ \gamma_0=\pa{\gamma_\rho,\rho,\vec{h}},\quad \lambda^0=\pa{-\vec{v}\cdot\vec{\lambda},\vec{\lambda}},\quad \vec{\lambda}=\vec{g}-\vec{\Omega}\times\vec{v},
$$
where $\gamma_\rho$ is an arbitrary function and $\vec{h}\cdot\vec{\lambda}=0$. Substituting (\ref{eq:4.3}.2) into equation (\ref{eq:4.1}) we find the inhomogeneous simple acoustic elements
\begin{aleq*}
A^0_\epsilon: &\gamma_0=\pa{\gamma_\rho,\frac{\kappa p}{\rho}\gamma_\rho,\frac{\rho^{\frac{1}{2}}}{\epsilon\rho (\kappa p)^{\frac{1}{2}}|\vl|}\pa{\rho\pa{\vec{g}-\vec{\Omega}\times\vec{v}-\frac{\kappa p}{\rho}\gamma_\rho\vl}}},\\
&\lambda^0=\pa{\epsilon\pa{\frac{\kappa p}{\rho}}^{\frac{1}{2}}|\vl|-\vec{v}\cdot\vec{\lambda}},\qquad \pa{\vec{g}-\vec{\Omega}\times\vec{v}}\cdot\vl=0.
\end{aleq*}%
The solution of the algebraic equations (\ref{eq:4.1}) under condition (\ref{eq:2.3}) leads to the inhomogeneous simple hydrodynamic elements
\begin{aleq}\label{eq:4.6}
H^0:\ &\gamma_0=\Bigg(\frac{-\rho^2\pa{\vec{g}-\vec{\Omega}\times\vec{v}}\cdot\vl}{\rho(\delta|\vl|)^2-\kappa p\vl^2},\frac{-\kappa\rho p\pa{\vec{g}-\vec{\Omega}\times\vec{v}}\cdot\vl}{\rho(\delta|\vl|)^2-\kappa p\vl^2},\\
 &\qquad\frac{1}{\delta|\vl|}\pa{\vec{g}-\vec{\Omega}\times\vec{v}+\frac{\kappa p\pa{\vec{g}-\vec{\Omega}\times\vec{v}}\cdot\vl}{\rho(\delta|\vl|)^2-\kappa p\vl^2}}\vec{\lambda}\Bigg),\\
&\lambda^0=\pa{\delta|\vl|-\vec{v}\cdot\vl,\vl},\ \text{where }\delta|\vl|\neq0,\ \epsilon\rho^{-1/2}(\kappa p)^{1/2}.
\end{aleq}%
\begin{center}
\ \hspace{1cm}\ 
\begin{table}
\fontsize{8}{8}\selectfont
\centering
\begin{tabular}{|l|l|l|}
\hline
type & solution & parameters\\
\hline
$\begin{aligned}&E^0\\ &\vec{g}\cdot\vec{\Omega}\neq0\end{aligned}$ & $\begin{aligned}&(\rho,p,\vec{v})=\pa{\dot{p}(r^0),p(r^0),\vec{v}_0},\\
&r^0(t,x)=-\vec{v}_0\cdot\vec{g}t+(\vec{g}-\vec{\Omega}\times\vec{v}_0)\cdot\vec{x},\\
&\vec{g}\neq \vec{\Omega}\times\vec{v}_0\end{aligned}$ & $\begin{aligned}&\text{constants: }\vec{v}_0;\\ &\text{functions: }p(r^0)>0,\\
&\dot{p}>0\end{aligned}$\\
\hline
$\begin{aligned}&E^0\\ &\vec{g}\cdot\vec{\Omega}=0\end{aligned}$
&$\begin{aligned}&(\rho,p,\vec{v})=\pa{\dot{p}(r^0),p(r^0),\nu(r^0)\vec{\Omega}+\vec{A}},\\
&r^0(t,x)=-\vec{A}\cdot\vec{g}t+(\vec{g}-\vec{\Omega}\times\vec{A})\cdot\vec{x},\\
&\vec{A}\cdot\vec{\Omega}=0,\ \vec{g}\neq\vec{\Omega}\times\vec{A}\end{aligned}$
&$\begin{aligned}
&\text{constants: }\vec{A};\\
&\text{functions: }p(r^0)>0,\\
&\nu(r^0);\ \dot{p}>0
\end{aligned}$\\
\hline
$\begin{aligned}&A^0_\epsilon\\ &\vec{\lambda}\times\vec{\Omega}\neq0\end{aligned}$
&$\begin{aligned}&\rho=\rho_0,\quad p=p_0,\quad\vec{v}(r^0)=\vec{g}\cdot(\vec{\Omega}\times\vec{c})\vec{c}\\
&+\pa{\epsilon \pa{\frac{\rho_0}{\kappa p_0}}^{\frac{1}{2}}\vec{g}\cdot\vec{\Omega}r^0+b_0}\vec{\Omega}+\vec{g}\cdot\vec{c}(\vec{c}\times\vec{\Omega}),\\
&\text{where }\\
&r^0(t,\vec{x})=\pa{\epsilon\pa{\frac{\kappa p_0}{\rho_0}}^{\frac{1}{2}}-\vec{g}\cdot(\vec{\Omega}\times\vec{c})}t+\vec{c}\cdot\vec{x},\\
&\vec{c}\cdot\vec{\Omega}=0,\ \vec{g}\cdot\vec{\Omega}\neq0\end{aligned}$
&$\begin{aligned}&\text{constants: }\vec{c},\ b_0,\\
&\rho_0>0,\ p_0>0;\\
&|\vec{c}|=1\end{aligned}$\\
\hline
$\begin{aligned}&A^0_\epsilon\\ &\vec{\lambda}=\epsilon_1|\vec{\lambda}|\vec{\Omega}\end{aligned}$
&$\begin{aligned}&\rho=\rho_0,\quad p=p_0,\quad \vec{v}(r^0)=A_1\cos\theta(r^0)\vec{g}\\
&+\epsilon_1\pa{\epsilon\pa{\frac{\kappa p_0}{\rho_0}}^{\frac{1}{2}}-c_0}\vec{\Omega}+\pa{1-A_1\sin\theta(r^0)}\vec{g}\times\vec{\Omega},\\
&\text{where }r^0(t,\vec{x})=c_0t+\epsilon\vec{\Omega}\cdot\vec{x},\quad\vec{g}\cdot\vec{\Omega}=0,\\
&\theta(r^0)=\epsilon \pa{\frac{\kappa p_0}{\rho_0}}^{-1/2}r^0+B_1\end{aligned}$
&$\begin{aligned}&\text{constants: }\rho_0>0,\\
&p_0>0\ A_1\neq 0,\\
&B_1,\ c_0\end{aligned}$\\
\hline
$\begin{aligned}&H^0\\ &\vec{\lambda}\times\vec{\Omega}\neq0,\\
&\lambda\cdot\vec{\Omega}=0\end{aligned}$
& $\begin{aligned}&\rho=\rho(r^0),\quad p=A\rho^\kappa,\quad\vec{v}(r^0)=\pa{\frac{1}{a_1\rho(r^0)}-c_0}\vec{c}\\
&+\pa{\vec{g}\cdot\vec{\Omega}a_1\int_0^{r^0}\rho(s)ds+b_1}\vec{\Omega}\\
&+\pa{\pa{\vec{g}\cdot(\vec{c}\times\vec{\Omega})}a_1\int_0^{r^0}\rho(s)ds+r^0+T_1}\vec{c}\times\vec{\Omega},\\
&\text{where }r^0(t,\vec{x})=c_0t+\vec{c}\cdot\vec{x},\\
&\text{and $\rho>0$ is given implicit form by:}\\
&r^0(t,x)=\pm\int_{\rho_0}^{\rho(r^0(t,x))}\pa{a_1^{-2}\rho^{-3}-\kappa A\rho^{\kappa-2}}\\
&\cdot\Bigg((T_1-\vec{g}\cdot\vec{c})^2-a_1^{-2}(\rho^{-2}-\rho_0^{-2})-F(\rho)\\
&-2(\vec{g}\cdot(\vec{c}\times\vec{\Omega}))\bigg(a_1A(\rho^\kappa-\rho_0^\kappa)a_1^{-1}(\rho^{-1}-\rho_0^{-1})\bigg)\Bigg)^{-\frac{1}{2}}d\rho,\\
&\text{where $\rho_0=\rho(0)$, $\rho\neq-\pa{a_1(\vec{g}\cdot(\vec{c}\times \vec{\Omega})-c_0)}^{-1}$ and}\\
&F(\rho)=\left\{\begin{aligned}&2\pa{\frac{\kappa A}{\kappa-1}}\pa{\rho^{\kappa-1}-\rho_0},\ &&\text{if }\kappa\neq1,\\
& 2A\ln\frac{\rho}{\rho_0}, &&\text{if }\kappa=1.\end{aligned}\right. \end{aligned}$
&$\begin{aligned}&\text{constants: }A>0,\\
&a_1\neq0,\ b_1,\ T_1\\
&\ c_0,\ \vec{c},\ |\vec{c}|=1;\\
& \vec{c}\cdot\vec{\Omega}=0,\\
&\vec{g}\cdot\vec{\Omega}\neq 0\end{aligned}$\\
\hline
$\begin{aligned}&H^0\\ &\vec{\lambda}=\epsilon_1|\vec{\lambda}|\vec{\Omega},\\
&\vec{g}\cdot\vec{\Omega}=0\end{aligned}$
&$\begin{aligned}&\rho=\rho_0,\quad p=p_0,\quad \vec{v}(r^0)=A_1\cos\theta(r^0)\vec{g}\\
&+B_0\vec{\Omega}+\pa{1-A_1\sin\theta(r^0)}\vec{g}\times\vec{\Omega},\\
&\text{where }\theta(r^0)=B_1+(c_0+\epsilon_1B_0)^{-1}r^0,\\
&r^0(t,\vec{x})=c_0t+\epsilon_1\vec{\Omega}\cdot\vec{x}\end{aligned}$
&$\begin{aligned}&\text{constants: }\rho_0>0,\\
&p_0>0,\ A_1\neq 0,\\
&B_1,\ c_0,\\
&B_0\neq -\epsilon_1c_0\end{aligned}$\\
\hline
$\begin{aligned}&H^0\\ &\vec{g}=\epsilon_2|\vec{g}|\vec{\Omega}\end{aligned}$
&$\begin{aligned}&\rho=\rho(r^0),\quad p=A\rho^\kappa,\quad \vec{v}=A_1\cos\theta(r^0)\vec{e}_1\\
&-A_1\sin\theta(r^0)\vec{e}_2+\pa{K\rho^{-1}-c_0}\vec{e}_3,\\
&\text{where }r^0(t,\vec{x})=c_0t+\epsilon_1\vec{e}_3\cdot\vec{x},\\
&\text{where $\rho>0$ is given implicitly by}\\
&r^0=\left\{\begin{aligned}&\frac{\epsilon_1\epsilon_2}{|\vec{g}|}\pa{\pa{\frac{\kappa A}{\kappa-1}}\rho^{\kappa-1}+\frac{K^2}{2\rho^2}}+\rho_0,\quad &&\kappa\neq 1,\\
&\frac{\epsilon_1\epsilon_2}{|\vec{g}|}\pa{A\ln\rho+\frac{K^2}{2\rho^2}}+\rho_0, &&\kappa=1\end{aligned}\right.\end{aligned}$
&$\begin{aligned}&\text{constants: }A>0,\\
&A_1,\ B_1, c_0,\\
&K\neq0,\ \rho_0,\\
&\vec{\Omega}=\vec{e}_3\end{aligned}$\\
\hline
\end{tabular}
\caption{Simple state solutions of the Euler system (\ref{eq:euler}). Here, we introduce the notation $E_0$ for the simple entropic state, $A^0_\epsilon$ for the simple acoustic state and $H^0$ for the simple hydrodynamic state.}
\end{table}
\end{center}
The rank-1 solutions, called simple states, are known \cite{Grundland:1974:b} and are summarized in Table 1. From the definition (\ref{eq:4.2}) it follows that the velocity of the entropic state $E^0$ relative to the fluid is equal to zero. This state propagates together with the fluid and not relative to it. The velocity of propagation of the acoustic state $A^0_\epsilon$ relative to the medium is equal to the sound velocity:
$$s^{1/2}=\epsilon\pa{dp/d\rho}^{1/2}.$$%
The sign $\epsilon=\pm1$ means that the state propagates in the left or in the right direction with respect to the medium. However, the hydrodynamic state $H^0$ can propagate relative to the medium with any speed except for the entropic velocity $\delta|\vl|=0$ and the acoustic velocity $\delta|\vl|=\epsilon|\vl|s^{1/2}$.
\paragraph{}In the further analysis of our considerations we will deal with an investigation of the homogeneous system (\ref{eq:euler}). We shall determine homogeneous simple elements which later enable us to construct some more general classes of solutions than the simple states. These solutions will represent the propagation of a single simple wave on a simple state. The algebraic equations which determine homogeneous simple elements for the equation (\ref{eq:euler}) are of the form
\begin{aleq}\label{eq:4.8}
&\rho\delta|\vl|\vec{\gamma}+\gamma_p\vec{\lambda}=0,\\
&\delta|\vl|\gamma_\rho+\rho\vec{\gamma}\cdot\vec{\lambda}=0,\\
&\delta|\vl|\pa{\rho\gamma_p-\kappa p\gamma_\rho}=0.
\end{aleq}%
System (\ref{eq:4.8}) is a linear homogeneous system with respect to the vector $\gamma=\pa{\gamma_\rho, \gamma_p, \vec{\gamma}}$. Consequently, the non-zero solution $\gamma$ exists if and only if the characteristic determinant of the system (\ref{eq:4.8}) vanishes
\begin{eqe}\label{eq:4.9}
(\delta|\vl|)^3\pa{(\delta|\vl|)^2-\frac{\kappa p}{\rho}|\vec{\lambda}|^2}=0.
\end{eqe}%
Equation (\ref{eq:4.9}) has two distinct types of solutions for the function $\delta|\vl|$. They are
\begin{aleq}\label{eq:4.10}
&1.\quad E:\ \delta_E=\delta|\vl|=0\text{ - entropic velocity}\\
&2.\quad A_\epsilon:\ \delta_{A_\epsilon}=\delta|\vl|=\epsilon|\vl|\sqrt{s}\text{ - sound velocity}.
\end{aleq}%
Equations (\ref{eq:4.8}) and (\ref{eq:4.10}.1) determine homogeneous entropic elements which are attributed to simple entropic waves
\begin{eqe}\label{eq:4.11}
E:\ \gamma_1^E=\pa{\gamma_\rho,0,\vec{h}},\quad \lambda^1_E=\pa{-\vec{v}\cdot\vec{\lambda}, \vec{\lambda}},
\end{eqe}%
where $\vec{h}\cdot\vec{\vl}=0$ and $\gamma_\rho$ is an arbitrary function. The equations (\ref{eq:4.8}), subjected to the condition (\ref{eq:4.10}.2), determine the homogenous acoustic elements
\begin{eqe}\label{eq:4.12}
A_\epsilon: \gamma_1^{A_\epsilon}=\pa{\gamma_\rho, s\gamma_\rho, -\epsilon \sqrt{s}\frac{\gamma_\rho}{\rho}\frac{\vec{\lambda}}{|\vl|}},\quad \lambda^1_{A_\epsilon}=\pa{\epsilon|\vl|\sqrt{s}-\vec{v}\cdot\vl, \vl},
\end{eqe}%
which correspond in turn to the simple acoustic waves $A_\epsilon$. Here $\gamma_\rho$ is an arbitrary function. The vector $\lambda^1$ in (\ref{eq:4.11}) or (\ref{eq:4.12}) can be treated as an analogue of a wave vector $(\omega,\vec{k})$ determining the velocity and the direction of the wave under consideration.
\section{Rank-2 solutions - wave propagation on a simple state}\label{sec:5}
\begin{figure}[h]
\begin{center}
\includegraphics[width=2.5in]{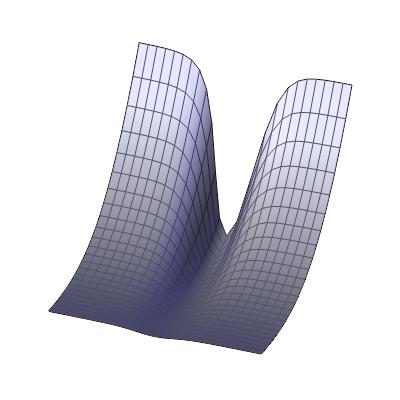}
\end{center}
\caption{Density as a function of $x$ and $z$ at time $t=0$ over the region $-5<x<5$, $0<x<1/2$. The values of parameters are given in (\ref{eq:E2.6}) and $K=-A^{1/2}, A=5/3$, $g=9.81$.}\label{fig:1}
\end{figure}

\begin{figure}[h]
\begin{center}
\includegraphics[width=2.5in]{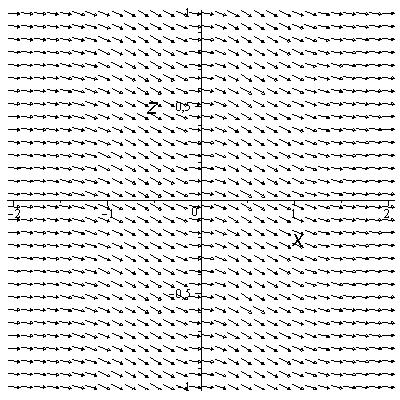}
\end{center}
\caption{Projection on the $xz$ plane of the flow of velocity. The values of the parameters are given in (\ref{eq:E2.6}) and $K=-A^{1/2}, A=5/3$.}\label{fig:2}
\end{figure}

Now, we present the rank-2 solutions of the initial equation (\ref{eq:euler}). These solutions are induced in the same way as a superposition of a simple wave on a simple state in the sense presented in Section \ref{sec:2} for which the conditions (\ref{eq:2.12}) hold. The propagations of a simple wave $E$ or $A_\epsilon$ on a simple state $E^0$, $A^0_\epsilon$ or $H^0$ are summarized in Table \ref{tab:2} and are denoted $E^0E$, $E^0A_\epsilon$, $A^0_\epsilon E$, $H^0E$ and $H^0A^0_\epsilon$. These solutions exist and are subject to certain conditions on the unknown functions which guarantee that they can be written in terms of Riemann invariants.
\begin{table}[b]
\begin{center}
\fontsize{14}{12} \selectfont
\begin{tabular}{|c|c|c|c|}
\hline
simple waves$\backslash$ simple states & $E$ &  $A^0_\epsilon$ & $H^0$\\
\hline $E$ &  $+$ & $+$ & $+$\\
\hline $A_\epsilon$ &  $+$ & $-$  & $+$ \\
\hline
\end{tabular}
\fontsize{10}{10} \selectfont
\caption{Summary of the superpositions of a simple wave on a simple state, where $+$ denotes that a superposition exists and $-$ that a rank-2 solution does not occur.}
\label{tab:2}
\end{center}
\end{table}
\paragraph{}Let us now illustrate our theoretical considerations of rank-2 solution from Section \ref{sec:2} by examples involving two separate cases, namely, an example when the coefficient $\alpha_1$ in (\ref{eq:2.12}) does not vanish anywhere and when $\alpha_1$ vanishes identically.
\paragraph{Example 1.} The study of superpositions of entropic waves $E$ on a simple state $E_0$ has revealed the existence of rank-2 solutions $E^0E$, written in terms of Riemann invariants. By choosing a proper length of the entropic vector fields $\gamma_0^{E^0}$ and $\gamma_1^{E^1}$ such that $\br{\gamma_0^{E^0},\gamma_1^{E^1}}=0$ holds, we look for a solution of the system of PDEs
\begin{aleq}\label{eq:E1.d}
&\frac{\del \rho}{\del r^0}=\gamma_{0\rho},\qquad &&\frac{\del \rho}{\del r^1}=0,\\
&\frac{\del p}{\del r^0}=\rho, &&\frac{\del p}{\del r^1}=0,\\
&\frac{\del \vec{v}}{\del r^0}=\vec{h}_0, &&\frac{\del \vec{v}}{\del r^1}=\vec{h}_1,
\end{aleq}%
where $\vec{h}_0, \vec{h}_1\in\RR^3$ are vector functions of $r^0$ and $r^1$ that satisfy the equations $\vec{h}_0\cdot\pa{\vec{g}-\vec{\Omega}\times\vec{v}}=0$ and $\vec{h}_1\cdot\vec{\lambda}^1=0$. A particular solution of this system in terms of $r^0$ and $r^1$ has the form
\begin{eqe}\label{eq:E1.1}
\rho=\dot{p}(r^0), \quad p=p(r^0),\quad \vec{v}=\pa{\beta(r^0,r^1),\frac{c_2+\Omega_2 \beta(r^0,r^1)}{\Omega_1},\frac{-c_1}{\abs{\vec{g}}^2}},
\end{eqe}%
where $\vec{\Omega}=\pa{\Omega_1,\Omega_2,0}$, $\vec{g}=(0,0,\abs{g}^2)$. Making use of the Proposition 2.2 for the case $\alpha_1=0$ and $\vec{g}\cdot\vec{\Omega}=0$, the Riemann invariants are determined by the following equations
\begin{aleq}\label{eq:E1.2}
&r^0=c_1t+\abs{\vec{g}}z,\quad c_1,c_2\in\RR,\\
&dr^1=\pa{c_1\alpha(r^0,r^1)-c_2}dt-\Omega_2dx+\Omega_1dy+\alpha(r^0,r^1)dz,
\end{aleq}%
where $\alpha$ and $\beta$ are arbitrary functions of $r^0$ and $r^1$, and $p$ is an arbitrary function of $r^0$. Now, choosing the constants $c_1=c_2=0$ and the function $\alpha=\alpha(r^1)$, and replacing them in the relation (\ref{eq:E1.2}) we obtain a definition of the Riemann invariants $r^0$ and $r^1$, where $r^1$ is an implicit function of $x$, $y$ and $z$
\begin{eqe}\label{eq:E1.3}
r^0=\abs{g}z,\quad r^1=\varphi(-\Omega_2x+\Omega_1y+\alpha(r^1)z),
\end{eqe}%
where $\varphi$ is an arbitrary function of its argument. Moreover, if we assume that $\varphi$ is the identity and $\alpha(r^1)=c(r^1+r^1_0)^{1/2}$, $c, r^1_0\in\RR$ then we can easily solve equation (\ref{eq:E1.3}) to obtain $r^1$ explicitly
\begin{eqe}\label{eq:E1.4}
r^1=-\Omega_2x+\Omega_1 y+\frac{1}{2}c^2 z^2\pm cz\pa{c^2z^2-\Omega_2 x+\Omega_1y-r_0^1}^{1/2}
\end{eqe}%
Various double Riemann wave solutions $E^0E$ can be found once the functions $\beta=\beta(r^0,r^1)$ and $p=p(r^0)$ are specified. By way of illustration we show how to obtain an eight-parameter family of explicit solutions with the freedom of one arbitrary function $a$ of $r^1$ by choosing
\begin{aleq}\label{eq:E1.5}
&p=\rho_0\arctan\pa{\sin(mr^0)}+p_0,\quad m, p_0, \rho_0, b\in\RR,\\
&\beta=\beta(r^0,r^1)=a(r^1)\operatorname{cn}\br{\pa{1+\cosh(br^0)}^{-1},k^2},
\end{aleq}%
where the modulus $k$ of the elliptic function $\operatorname{cn}$ satisfies $0<k<1$. Under the above assumptions, we exclude the possibility of a gradient catastrophe and we obtain the explicit double-wave solution $E^0E$
\begin{aleq}\label{eq:E1.6}
&\rho=\rho_0\cosh(m\abs{\vec{g}}z)\pa{1+\sinh^2(m\abs{g}z)}^{-1},\\
&p=\rho_0\arctan\pa{\sinh(m)\abs{\vec{g}}z}+p_0,\\
&\vec{v}=a(r^1)\operatorname{cn}\br{\pa{1+\cosh\pa{\arctan(b\abs{\vec{g}}z)}}^{-1},k^2}\pa{1,\Omega_2/\Omega_1,0},
\end{aleq}%
where $r^1$ is given explicitly in terms of $x$, $y$ and $z$ by the expression (\ref{eq:E1.4}). Obviously, other choices of functions $\alpha$, $\beta$ and $p$ lead to different double-wave solutions. The problem of the classification of these solutions still remains open. Nevertheless some results are known \cite{GrundlandZelazny:1983}. The stationary solution (\ref{eq:E1.6}) describes the compressible fluid in a state of equilibrium in the presence of gravitational and Coriolis forces. Note that the simple state $E^0$ interacts with the simple wave $E$, since the wave $E$ does not influence the state $E^0$. Therefore, according to \cite{Jeffrey:1976}, they interact independently.
\paragraph{Example 2.}Consider rank-2 solutions of type $E^0A_\epsilon$ which represent the propagation of a simple acoustic wave $A_\epsilon$ on simple entropic state $E^0$. For the case $\alpha_1=0$ and $\vec{g}\cdot\vec{\Omega}=0$, we choose the proper length of the entropic vector field $\gamma_0^{E^0}$ and the acoustic vector field $\gamma_1^{A_\epsilon}$ for which the commutator of these fields vanish (\textit{i.e.} $\br{\gamma_0^{E^0},\gamma_1^{A_\epsilon}}=0$). This requires the solving of the system of PDEs
\begin{aleq}\label{eq:E2.b}
&\frac{\del \rho}{\del r^0}=\gamma_{0\rho},\qquad &&\frac{\del \rho}{\del r^1}=\gamma_{1\rho},\\
&\frac{\del p}{\del r^0}=\rho, &&\frac{\del p}{\del r^1}=\frac{\kappa p}{\rho}\gamma_{1\rho},\\
&\frac{\del \vec{v}}{\del r^0}=\vec{h}_0, &&\frac{\del \vec{v}}{\del r^1}=-\epsilon\pa{\frac{\kappa p}{\rho}}^{1/2}\frac{\gamma_{1\rho}}{\rho}\frac{\vec{\lambda}^1}{\abs{\vec{\lambda}^1}}.
\end{aleq}%
From the integration of (\ref{eq:E2.b}) we find a particular solution in a parametric form in terms of $r^0$ and $r^1$
\begin{gaeq}\label{eq:E2.c}
\rho=\exp\pa{\frac{r^0}{A}+B(r^1)},\qquad p=A\rho,\\
\vec{v}=-e^{\phi(r^0)}\frac{c_0}{|\vec{g}|^2}\vec{g}-\pa{B_0+\epsilon\epsilon_1 A^{1/2}B(r^1)}\vec{\Omega}+\pa{1\mp e^{\phi(r^0)}\frac{(\abs{\vec{g}}^2-c_0^2)^{1/2}}{\abs{\vec{g}}^2}}\vec{g}\times\vec{\Omega}
\end{gaeq}%
The procedure, as described in Proposition 2.2 of Section 2, allows us to obtain the Riemann invariants $r^0$ and $r^1$ implicitly defined by the equations
\begin{gaeq}\label{eq:E2.d}
c_0t\pm\frac{\pa{\abs{\vec{g}}^2-c_0^2}^{1/2}}{\abs{\vec{g}}^2}\vec{g}\cdot\vec{x}-\frac{c_0}{\abs{\vec{g}}^2}\pa{\vec{g}\times\vec{\Omega}}\cdot\vec{x}+a_0=\int_0^{r^0}e^{-\phi(\xi)}d\xi,\\
\psi(r^1)=\pa{\epsilon\epsilon_1 A^{1/2}(B(r^1)+1)+B_0}t+\vec{\Omega}\cdot\vec{x},
\end{gaeq}%
where $A>0$, $B_0$, $c_0$ and $a_0$ are arbitrary constants, and $\phi(r^0)$, $B(r^1)$ and $\psi(r^1)$ are arbitrary functions. For the solution to be of rank 2, it is necessary that $\dot{B}(r^1)\neq0$. If we choose the arbitrary function $F(r^0)$ to be
\begin{eqe}\label{eq:E2.1}
F(r^0)=F_0\arctanh\sqrt{\exp(r^0/A_0)-1},
\end{eqe}%
then this particular choice allows us to determine the invariant $r^0$ in the explicit form
\begin{eqe}\label{eq:E2.2}
r^0=A_0\ln\pa{\tanh^2(F_0\zeta)+1},
\end{eqe}%
where
\begin{eqe}\label{eq:E2.3}
\zeta=c_0t+\frac{\epsilon_2\sqrt{|g|^2-c_0^2}}{|g|^2}\vec{g}\cdot\vec{x}-\frac{c_0(\vec{g}\times\vec{\Omega})\cdot\vec{x}}{|g|^2}+c_1.
\end{eqe}%
The function $\Psi(r^1)$ is chosen so as to be proportional to $B(r^1)$, that is
\begin{eqe}\label{eq:E2.4}
\Psi(r^1)=KB(r^1),\quad K\in\RR.
\end{eqe}%
In this case, the solution $B(r^1)$ becomes
\begin{eqe}\label{eq:E2.5}
B(r^1)=\frac{\pa{\epsilon\epsilon_1\sqrt{A}+B_0}t+\vec{g}\cdot\vec{x}}{K-\epsilon\epsilon_1\sqrt{A}t}.
\end{eqe}%
With the choice of parameters
\begin{gaeq}\label{eq:E2.6}
\epsilon=\epsilon_1=1,\quad B_0=0,\quad F_0=1,\quad A_0=5,\\
\vec{\Omega}=(0,-1,0),\quad c_0=g,\quad c_1=0,\quad \vec{g}=(0,0,g),
\end{gaeq}%
the non-stationary solution $E^0A_\epsilon$ of rank 2 takes the form
\begin{gaeq}\label{eq:E2.7}
\rho=\pa{\tanh^2(gt-x)+1}^{5/A}\exp\pa{\frac{A^{1/2}t+gz}{K-A^{1/2}t}},\qquad p=A \rho,\\
v_1=g,\qquad v_2=-\pa{\frac{At+A^{1/2}gz}{K-At}},\qquad v_3=-\frac{2\tanh^2(gt-x)}{2\cosh^2(gt-x)-1},
\end{gaeq}%
where $v_i$, $i=1,2,3$, are the components of the velocity. If the constant $K$ in expression (\ref{eq:E2.7}) is positive, then the gradient catastrophe occurs for the solution at time $t=K/A^{1/2}$. If $K$ is negative, the gradient catastrophe does not occur. However, in the case where $K<0$, the argument of the exponent in the density (\ref{eq:E2.7}) is negative, which represents damping. If the remaining parameters are given the values $K=-A^{1/2}, A=5/3$, the density $\rho$ can be visualized at a fixed time $t=0$ as a function at $x$ and $z$ (see fig. \ref{fig:1}).
 The factor $(\tanh^2(\zeta)+1)^{3/5}$ appearing in the density represents a kink solution which is bounded for the damping exponential. This kink wave propagates at the constant velocity $1/g$ in the positive $X$ direction. Under the hypotheses of (\ref{eq:E2.6}), the first component $v_1$ of the velocity is constant and the second component $v_2$ depends on $r^1$ via the function $B(r^1)$ given in (\ref{eq:E2.5}). The third component $v_3$ depends on $r^0$ only. The velocity field is represented in fig. \ref{fig:2}
 and propagates in the positive $x$ direction with constant velocity. The propagation of the acoustic simple wave $A_\epsilon$ on the entropic simple state $E_0$ describes a free fall of fluid in the field of gravitation subjected to the Coriolis force in which the given wave propagates.
\paragraph{Example 3.} Entropic wave $E$ on hydrodynamic state $H^0$ when $\alpha_1=0$. The propagation of the entropic wave $E$ admitted by the inhomogeneous system (\ref{eq:euler}) is determined by the following system of PDEs (\ref{eq:2.8})
\begin{aleq}\label{eq:E.1}
&(1)\  \frac{\del \rho}{\del r^0}=-\rho\frac{(\vec{g}-\vec{\Omega}\times \vec{v})\cdot\vec{\lambda}^0}{\delta^2-\frac{\kappa p}{\rho}|\vec{\lambda^0}|^2}, &&(4)\  \frac{\del \rho}{\del r^1}=\gamma_{\rho},\\
&(2)\  \frac{\del p}{\del r^0}=-\kappa\rho\frac{(\vec{g}-\vec{\Omega}\times \vec{v})\cdot\vec{\lambda}^0}{\delta^2-\frac{\kappa p}{\rho}|\vec{\lambda^0}|^2},&&(5)\  \frac{\del p}{\del r^1}=0,\\
&(3)\  \frac{\del\vec{v}}{\del r^0}=\frac{1}{\delta\rho}\pa{\rho\pa{\vec{g}-\vec{\Omega}\times\vec{v}}+\kappa p\frac{(\vec{g}-\vec{\Omega}\times \vec{v})\cdot\vec{\lambda}^0}{\delta^2-\frac{\kappa p}{\rho}|\vec{\lambda^0}|^2}\vec{\lambda}^0,}&&(6)\  \frac{\del \vec{v}}{\del r^1}=\vec{h},\\
\end{aleq}%
where $\vec{h}\cdot\vec{\lambda}^1=0$ while $\vec{\lambda}^0$ and $\vec{\lambda}^1$ are the solutions
\begin{eqe}\label{eq:E.2}
\lambda^0=C e^{\phi(r^0)},\quad \lambda^1=\mu(r)\pa{\tilde{C\chi(r)}+Y(r^1)},\quad C\in\RR^4
\end{eqe}%
of the system (\ref{eq:2.12}) when $\alpha_1=0$ and must satisfy the constraints (\ref{eq:4.6}) and (\ref{eq:4.11}), namely
\begin{eqe}\label{eq:E.3}
\lambda^0=\pa{\delta-\vec{v}\cdot\vec{\lambda}^0,\vec{\lambda}^0},\qquad \lambda^1=\pa{-\vec{v}\cdot\vec{\lambda}^1,\vec{\lambda}^1}.
\end{eqe}%
For the sake of simplicity, we set
$$c=\pa{c_0,\vec{c}}=\pa{\frac{C_0}{|\vec{C}|}, \frac{\vec{C}}{|\vec{C}|}},\quad \Phi(r^0)=\phi(r^0)+ln|\vec{C}|,\quad \chi(r)=|\vec{C}|\tilde{\chi}(r),$$%
where $C=\pa{C_0,\vec{C}}$, so we can rewrite the solutions (\ref{eq:E.2}) for $\lambda^0$ and $\lambda^1$ in the form
\begin{eqe}\label{eq:E.4}
\lambda^0=ce^{\Phi(r^0)},\qquad \lambda^1=\mu(r)\pa{\chi(r)c+Y(r)},
\end{eqe}%
with $c=\pa{c_0,\vec{c}}$ and $|\vec{c}|=1$. As a consequence of the equations (\ref{eq:E.3}) and (\ref{eq:E.4}), we may write
$$
\vec{\lambda}^0=\vec{c}e^{\Phi(r^0)},\qquad \delta=\pa{c_0+\vec{v}\cdot\vec{c}}e^{\Phi(r^0)}.
$$
\paragraph{}Now, we proceed to solve the system (\ref{eq:E.1}) under the assumption that $\vec{\lambda^0}\cdot\vec{\Omega}=e^{-\Phi(r^0)}\vec{c}\cdot\vec{\Omega}=0$. We consider here the case when $\vec{c}\times\vec{\Omega}\neq 0$, so we can express the vectors $\vec{v}$ and $\vec{g}$ in the orthonormal basis $\ac{\vec{c},\vec{\Omega},\vec{c}\times\vec{\Omega}}$ assuming $|\vec{\Omega}|=1$. Then we set
\begin{eqe}\label{eq:E.6}
\vec{v}=v_1(r)\vec{c}+v_2(r)\vec{\Omega}+v_3(r)\vec{c}\times\vec{\Omega},\qquad \vec{g}=g_1\vec{c}+g_2\vec{\Omega}+g_3\vec{c}\times\vec{\Omega},
\end{eqe}%
where the components $g_1$, $g_2$ and $g_3$ of the gravitational vector are constants. From equation (\ref{eq:E.1}.5), it is clear that the pressure $p$ is a function of $r^0$ only. Considering this fact when comparing equations (\ref{eq:E.1}.1) and (\ref{eq:E.1}.2) leads to the PDE $\rho^{-1}\del\rho/\del r^0=(\kappa p)^{-1}\del p/\del r^0$ which is easily solved to obtain the density in terms of the pressure and an arbitrary function $A$ of $r^1$ in the form
\begin{eqe}\label{eq:E.7}
\rho=\pa{\frac{p(r^0)}{A(r^1)}}^{1/\kappa}.
\end{eqe}%
Now, considering (\ref{eq:E.7}), the equations (\ref{eq:E.1}.1) and (\ref{eq:E.1}.2) are equivalent. So, using the hypothesis that $\vec{c}\cdot\vec{\Omega}=0$ and introducing the vector field and the gravity vector given by (\ref{eq:E.6}) into equations (\ref{eq:E.1}.2) (or  (\ref{eq:E.1}.1)) and  (\ref{eq:E.1}.3) gives, after comparing the components in the vectorial equation, the system of PDEs
\begin{aleq}\label{eq:E.8}
&\frac{\del p}{\del r^0}=-\kappa pe^{-\Phi(r^0)}\frac{g_1-v_3}{(c_0+v_1)^2-\kappa A^{1/\kappa}(r^1)p^{1-1/\kappa}}\\
&\frac{\del v_1}{\del r^0}=e^{-\Phi(r^0)}\frac{\pa{g_1-v_3}\pa{c_0+v_1}}{(c_0+v_1)^2-\kappa A^{1/\kappa}(r^1)p^{1-1/\kappa}},\\
&\frac{\del v_2}{\del r^0}=e^{-\Phi(r^0)}\frac{g_2}{c_0+v_1},\\
&\frac{\del v_3}{\del r^0}=e^{-\Phi(r^0)}\pa{1+\frac{g_3-c_0}{c_0+v_1}}.
\end{aleq}%
By simple quadratures of the two last equations in (\ref{eq:E.8}), we find that
\begin{aleq}\label{eq:E:9}
&v_2=g_2\int_0^{r^0}\frac{e^{-\Phi(\xi)}}{v_1(\xi,r^1)+c_0}d\xi+V_2(r^1),\\
&v_3=(g_3-c_0)\int_0^{r^0}\frac{e^{-\Phi(\xi)}}{v_1(\xi,r^1)+c_0}d\xi+\int_0^{r^0}e^{-\Phi(\xi)}d\xi+V_3(r^1),
\end{aleq}%
where $V_2(r^1)$ and $V_3(r^1)$ are arbitrary functions of integration. Comparing the two first equations in (\ref{eq:E.8}) yields
$$-\frac{1}{\kappa p}\frac{dp}{dr^0}=\frac{1}{v_1+c_0}\frac{\del v_1}{\del r^0},$$%
which has the solution
\begin{eqe}\label{eq:E.10}
c_0+v_1=\frac{1}{p^{1/\kappa}S(r^1)},
\end{eqe}%
where $S(r^1)$ is an arbitrary function of integration. Hence, using (\ref{eq:E.10}), we can eliminate the quantity $c_0+v_1$ from equation (\ref{eq:E.8}.1) which becomes
\begin{eqe}\label{eq:E.11}
\pa{\frac{p^{-1-1/\kappa}}{\kappa S(r^1)^2}-A(r^1)^{1/\kappa}}p(r^0)^{-1/\kappa}\dot{p}(r^0)=e^{-\Phi(r^0)}(v_3-g_1).
\end{eqe}%
\paragraph{}We now proceed to show that equation (\ref{eq:E.11}) has no solution when $\dot{S}(r^1)\neq0$ and $\dot{V}_3(r^1)\neq0$. If the pressure $p=p_0$ is constant then the left hand side of equation (\ref{eq:E.11}) vanishes, so the same happens to the right hand side. Next, we take the derivative with respect to $r^0$. Since the result must vanish, this implies that $S(r^1)=-(g_3-c_0)^{-1}p_0^{-\kappa/1}$ and we conclude that $S$ is constant. If $\dot{p}(r^0)\neq 0$, we first rewrite equation (\ref{eq:E.11}) in the form
\begin{eqe}\label{eq:E.12}
v_3=g_1+e^{\Phi(r^0)}\pa{\frac{p^{-1-1/\kappa}}{\kappa S(r^1)^2}-A(r^1)^{1/\kappa}}p(r^0)^{-1/\kappa}\dot{p}(r^0).
\end{eqe}%
Next, we introduce $v_3$, given by (\ref{eq:E.12}), into the last equation of the system (\ref{eq:E.8}) and then we take the derivative with respect to $r^1$. Dividing the result by $3(g_3-c_0)S^2$, we find that
\begin{eqe}\label{eq:E.13}
p^{1/\kappa}(r^0)\dot{S}(r^1)=\frac{1}{3(g_3-c_0)S^2(r^1)}\frac{d}{dr^1}\pa{A^{1/\kappa}(r^1)S^2(r^1)}.
\end{eqe}%
Differentiating again, now with respect to $r^0$, leads to $\frac{1}{\kappa}\dot{p}p^{-1+1/\kappa}\dot{S}=0$. Since we assumed that $\dot{p}\neq 0$, the only possibility is that $\dot{S}=0$ and so $S$ is constant. Consequently, equation (\ref{eq:E.13}) implies that we also have $\dot{A}=0$ in the case where $\dot{p}\neq 0$. For the rest of the example, we restrict ourselves to the case when $\dot{p}\neq0$. Now, we suppose that $A$ and $S=S_1$ are constant. Under these hypotheses, we deduce from equation (\ref{eq:E.12}) that $v_3$ is a function of $r^0$ only so $V_3(r^1)=T_1$ is a constant. Equations (\ref{eq:E.7}), (\ref{eq:E:9}) and (\ref{eq:E.10}) now take the form
\begin{aleq}\label{eq:E.14}
&p=p(r^0),\\
&\rho=A^{-1/\kappa}p(r^0)^{1/\kappa},\\
&v_1=\frac{1}{p(r^0)^{1/\kappa}S_1}-c_0,\\
&v_2=g_2\int_0^{r^0}e^{-\Phi(\xi)} p(r^0)^{1/\kappa}S_1d\xi+V_2(r^1),\\
&v_3=(g_3-c_0)\int_0^{r^0}e^{-\Phi(\xi)}p(r^0)^{1/\kappa}S_1d\xi+\int_0^{r^0}e^{-\Phi(\xi)}d\xi+T_1.
\end{aleq}%
At this point, $p$, $\rho$, $v_1$ and $v_3$ depend on $r^0$ only, so we make the hypothesis $\dot{V}_2\neq0$ otherwise the final solution will be of rank one and no superposition will occur. Now, introducing the solution (\ref{eq:E.4}) for $\lambda^1$ into the second equation in (\ref{eq:E.3}), we obtain
\begin{aleq}\label{eq:E.15}
&(1) \quad &&-\vec{v}\cdot\vec{\lambda}^1=\mu(r)\pa{\chi(r)c_0+Y_0(r^1)},\\
&(2) &&\vec{\lambda}^1=\mu(r)\pa{\chi(r)\vec{c}+\vec{Y}(r^1)}.
\end{aleq}%
Taking the scalar product of the equation (\ref{eq:E.1}.6) with the vector $\vec{\lambda}^1$ yields $(\del\vec{v}/\del r^1)\cdot\vec{\lambda}^1=0$ in which we substitute the value of $\vec{\lambda}^1$ given by (\ref{eq:E.15}.2) and $\vec{v}$ given by (\ref{eq:E.6}) to obtain the condition $\mu(r)\dot{V}_2(r^1)\vec{\Omega}\cdot\pa{\chi(r)\vec{c}+\vec{Y}(r^1)}=0$. Setting $\vec{Y}=Y_1\vec{c}+Y_2\vec{\Omega}+Y_3\vec{c}\times\vec{\Omega}$ and using the fact that $\vec{\Omega}\cdot\vec{c}=0$, this last condition takes the form $\dot{V}_2Y_2=0$. Since we have assumed earlier that $\dot{V}_2\neq0$, we conclude that $Y_2=0$. Considering this result, we can solve the equation (\ref{eq:E.15}.1) to obtain $\chi(r)$ given by
\begin{aleq*}
\chi(r)=&-p^{1/\kappa} S_1\pa{-c_0Y_1+Y_0+T_1Y_3}-\pa{g_3-c_0}S_1^2p^{1/\kappa}\int_0^{r^0}p(\xi)^{1/\kappa}e^{-\Phi(\xi)}d\xi\\
&-p^{1/\kappa}S_1Y_3\int_0^{r^0}e^{-\Phi(\xi)}d\xi-Y_1,
\end{aleq*}%
where we have introduced the solution for $v_1$, $v_2$ and $v_3$ given by (\ref{eq:E.14}). Using the fact that $\int_0^{r^0}e^{-\Phi(\xi)}p^{1/\kappa}\pa{\int_0^{\xi}e^{-\Phi(\eta)}p^{1/\kappa}d\eta}d\xi=\frac{1}{2}\pa{\int_0^{r^0}p^{1/\kappa}e^{-\Phi(\xi)}d\xi}^2$, we can find the value of the integral $\int_0^{r^0}\chi(\xi,r^1)e^{-\Phi(\xi)}d\xi$ which enables us to find the Riemann invariants in the form
\begin{aleq}\label{eq:E.16}
&\int_0^{r^0}e^{-\Phi(\xi)}d\xi=c_0t+\vec{c}\cdot\vec{x}+a_0,\\
&-\pa{Y_0-c_0Y_1+T_1Y_3}S_1\int_0^{r^0}e^{-\Phi(\xi)}p(\xi)^{1/\kappa}d\xi+\pa{Y_0-c_0Y_1}t\\
&+\pa{\vec{Y}-Y_0\vec{c}}\cdot\vec{x}-Y_3S_1\int_0^{r^0}p(\xi)^{1/\kappa}e^{-\Phi(\xi)}\pa{\int_0^\xi e^{-\Phi(\eta)}d\eta}d\xi\\
&-\frac{g_3-c_0}{2}S_1^2\pa{\int_0^{r^0}p(\xi)^{1/\kappa}e^{-\Phi(\xi)}d\xi}^2=\psi(r^1),
\end{aleq}%
where the quantities $p$ and $\Phi$ must satisfy the equation
\begin{aleq}\label{eq:E.17}
&G(p(r^0))\dot{p}(r^0)=l_1e^{-\Phi(r^0)}\\
&+e^{-\Phi(r^0)}l_2\int_0^{r^0}p(\xi)^{1/\kappa}e^{-\Phi(\xi)}d\xi+e^{-\Phi(r^0)}\int_0^{r^0}e^{-\Phi(\xi)}d\xi
\end{aleq}%
obtained by the substitution of the solution (\ref{eq:E.14}) into the PDE (\ref{eq:E.8}.1) and where we denote
$$G(p)=\pa{\frac{p^{-(1+1/\kappa)}}{\kappa S_1^2}-A^{1/\kappa}} p^{-1/\kappa},\quad l_1=T_1-g_1,\quad l_2=(g_3-c_0)S_1.$$%
The equation (\ref{eq:E.17}) is solved to find
\begin{aleq}\label{eq:E.18}
&e^{-\Phi(r^0)}=\pm \frac{G(p(\xi))\dot{p}(\xi)}{\bigg(l_1^2+2\int_0^{r^0}G(p(\xi))\dot{p}(1+l_2 p^{1/\kappa}(\xi))d\xi\bigg)^{1/2}}.
\end{aleq}%
So, integrating each side of equation (\ref{eq:E.18}) from 0 to $r^0$, we find the relation
\begin{eqe}\label{eq:E.19}
c_0t+\vec{c}\cdot\vec{x}+a_0=\pm\int_0^{r^0}\frac{G(p(\xi))\dot{p}(\xi)}{\bigg(l_1^2+2\int_0^{r^0}G(p(\xi))\dot{p}(1+l_2 p^{1/\kappa}(\xi))d\xi\bigg)^{1/2}}d\xi
\end{eqe}%
that defines the pressure $p$ in terms of the independent variables through the quantity $c_0t+\vec{c}\cdot\vec{x}+a_0$. In some sense, this relation means that the pressure itself plays the role of the Riemann invariant associated with the non-homogeneous wave vector $\lambda^0$. Making use of condition (\ref{eq:E.17}) and equation (\ref{eq:E.18}), the relation defining implicitly the Riemann invariant $r^1$ takes the form
\begin{aleq}\label{eq:E.20}
&\psi_1(r^1)=-Y_1(r^1)(c_0 t+\vec{c}\cdot \vec{x}+a_0)+Y_1(r^1)\vec{c}\cdot\vec{x}+Y_3(r^1)(\vec{c}\times\vec{\Omega})\cdot\vec{x}\\
&\mp S_1(Y_0(r^1)+g_1Y_3(r^1))\int_0^{r^0}\frac{p^{1/\kappa}G(p(\xi))\dot{p}(\xi)d\xi}{\pa{l_1^2+\int_0^{\xi}G(p(\eta))\dot{p}(\eta)(1+l_2p(\eta)^{\frac{1}{\kappa}})d\eta}^{\frac{1}{2}}}\\
&+Y_3(r^1)\int_0^{r^0}p(\xi)^{1/\kappa}G(p(\xi))\dot{p}(\xi)d\xi.
\end{aleq}%
\paragraph{}In summary, the solution (\ref{eq:E.14}) now takes the form
 \begin{aleq}\label{eq:E.23}
p&=p(t,\vec{x}),\\
\rho&=A^{-1/\kappa}p(t,\vec{x})^{1/\kappa},\\
v_1&=\frac{1}{p(t,\vec{x})^{1/\kappa}S_1}-c_0,\\
v_2&=V_2(r^1(t,x))+\vec{g}\cdot\vec{\Omega}\int_0^{r^0}\frac{p^{1/\kappa}G(p(\xi))\dot{p}(\xi)d\xi}{\pa{l_1^2+\int_0^{\xi}G(p(\eta))\dot{p}(\eta)(1+l_2p(\eta)^{1/\kappa})d\eta}^{1/2}}\\
v_3&=l_1+\bigg(l_1^2+2\int_0^{r^0}G(p(\xi))\dot{p}(1+l_2 p^{1/\kappa}(\xi))d\xi\bigg)^{1/2},
\end{aleq}%
where the quantities $v_1$, $v_2$ and $v_3$ are the components of the vector field $\vec{v}=v_1\vec{c}+v_2\vec{\Omega}+v_3\vec{c}\times\vec{\Omega}$ of the velocity of the fluid, $V_2(r^1)$ is an arbitrary function, $r^1(t,x)$ is defined implicitly by equation (\ref{eq:E.20}) and $p(t,\vec{x})$ is defined by (\ref{eq:E.19}) in which we replace $p(r^0)$ by $p(t,\vec{x})$ after computation of the integrals.
\paragraph{}In the particular case where $l_2=g_3-c_0=0$, and if we choose a coordinate system such that $\vec{x}=x\vec{c}+y\vec{\Omega}+z\vec{c}\times\vec{\Omega}$, the solution (\ref{eq:E.23}) becomes
\begin{aleq}\label{eq:E.24}
&\rho=A^{1/\kappa}p^{1/\kappa},\\
&v_1=\frac{1}{S_1p^{1/\kappa}}-c_0,\\
&v_2=V_2(r^1)\pm\int^{p(t,x)}\frac{\pa{p^{-1-1/\kappa}-A^{1/\kappa}}p^{-1/\kappa}}{\pa{l_1^2+(2S_1^2)^{-1}p^{-2/\kappa}-A^{1/\kappa}F(p)-a_1}^{1/2}}dp,\\
&v_3=c_0t+x+a_0+l_1+g_1,
\end{aleq}%
where $a_1$ is a constant of integration. The pressure $p$ is implicitly defined by the equation
\begin{eqe}\label{eq:E.25}
c_0t+x+a_0=\pm\pa{l_1^2+(2S_1^2)^{-1}p^{-2/\kappa}-A^{1/\kappa}F(p)-a_1}^{1/2},\\
\end{eqe}%
\begin{aleq*}
F(p)=\left\{\begin{aligned}&\ln p,\qquad &&\kappa=1,\\ &\frac{1}{1-1/\kappa}p^{1-1/\kappa}, &&\kappa\neq1,\end{aligned}\right.
\end{aleq*}
and the Riemann invariant $r_1$ is determined implicitly by the equation
\begin{aleq*}
\psi_1(r^1)&=\mp S_1(Y_0(r^1)+g_1Y_3(r^1))\int\frac{p^{1/k}\pa{\frac{p^{-1-1/\kappa}}{\kappa S_1^2}-A^{1/\kappa}}}{\pa{l_1^2+\frac{p^{-2/\kappa}}{2S_1^2}-A^{1/\kappa }F(p)}^{1/2}}dp\\
&-Y_1(r^1)(c_0t+x+a_0)+Y_3(r^1)\pa{\frac{\ln(p)}{\kappa S_1^2}-\frac{A^{1/\kappa}}{1+1/\kappa}p^{1+1/\kappa}}\\
&+xY_1(r^1)+zY_3(r^1).
\end{aleq*}%
The solution (\ref{eq:E.24}) represents a superposition of a simple wave parameterized by $r^1$ and a simple state parameterized by the pressure $p$ which plays the role of the invariant $r^0$. The simple state propagates at a phase velocity $g_3$ which is oriented along the $x$-axis. The simple wave interacts with the simple state and propagates in the $xz$-plane in the direction of $\vec{\lambda}^1$ which can be expressed in terms of the vectors $\vec{Y}(r^1)$. The phase velocity of  this wave is defined by $\lambda_0^1/|\vec{\lambda}^1|$. It should be noted that the second component of the velocity field is oriented along the vector $\vec{\Omega}$ corresponding to the Coriolis force. The simple wave associated with the wave vector $\vec{\lambda^1}$ only affects the component of the flow oriented along the $y$-axis. However, the vector $\vec{\lambda^1}$ lies in the $xz$-plane, which allows us to interpret the simple wave as a transversal wave. It should be noted that the function $V_2(r^1)$ is arbitrary, which implies that one can choose the profile of the wave associated with $\vec{\lambda^1}$ as required.
\paragraph{}Finally, we observe that for certain particular values of the polytropic coefficient, the pressure can be expressed in an explicit way. For example, for $\kappa=3$, the quantity $p^{2/3}$ appears in equation (\ref{eq:E.25}) in the form of an expression
$$p(t,x)=\pa{-A^{-1/3}\zeta(t,x)+\pa{\frac{A^{-2/3}}{9}\zeta(t,x)^2+\frac{A^{-1/3}}{3S_1^2}}^{1/2}}^{3/2},$$
where $\zeta(t,x)=a_1-l_1^2+c_0t+x+a_0$ and where we have considered only the positive branch since it is associated with a real value of p(t,x).
\paragraph{Example 4.} Entropic wave $A_\epsilon$ on hydrodynamic state $H^0$ when $\alpha_1\neq0$. The propagation of the acoustic wave $A_\epsilon$ on a hydrodynamic state $H^0$ admitted by the inhomogeneous system (\ref{eq:euler}) is determined by the following system of PDEs
\begin{aleq}\label{eq:A.1}
&(1)\  \frac{\del \rho}{\del r^0}=-\rho\frac{(\vec{g}-\vec{\Omega}\times \vec{v})\cdot\vec{\lambda}^0}{\delta^2-\frac{\kappa p}{\rho}|\vec{\lambda^0}|^2},\qquad (2)\  \frac{\del p}{\del r^0}=-\kappa\rho\frac{(\vec{g}-\vec{\Omega}\times \vec{v})\cdot\vec{\lambda}^0}{\delta^2-\frac{\kappa p}{\rho}|\vec{\lambda^0}|^2},\\
&(3)\  \frac{\del\vec{v}}{\del r^0}=\frac{1}{\delta\rho}\pa{\rho\pa{\vec{g}-\vec{\Omega}\times\vec{v}}+\kappa p\frac{(\vec{g}-\vec{\Omega}\times \vec{v})\cdot\vec{\lambda}^0}{\delta^2-\frac{\kappa p}{\rho}|\vec{\lambda^0}|^2}\vec{\lambda}^0},\\
&(4)\ \frac{\del \rho}{\del r^1}=\gamma_\rho,\qquad (5)\ \frac{\del p}{\del r^1}=\frac{\kappa p}{\rho}\gamma_\rho,\qquad (6)\ \frac{\del \vec{v}}{\del r^1}=-\epsilon \pa{\frac{\kappa p}{\rho}}^{1/2}\frac{\gamma_\rho}{\rho}\frac{\vec{\lambda}^1}{|\vec{\lambda^1}|},
\end{aleq}%
where $\gamma_\rho$ is an arbitrary function of $r=(r^0,r^1)$ while $\vec{\lambda}^0$ and $\vec{\lambda}^1$ are the solutions
\begin{eqe}\label{eq:A.2}
\lambda^0=\lambda(r^1) e^{\phi(r)},\quad \lambda^1=\tilde{\mu}(r)\pa{\frac{\del \phi}{\del r^1}\lambda(r^1)+\dot{\lambda}(r^1)},\quad \lambda(r^1)\in\RR^4,
\end{eqe}%
of the system (\ref{eq:2.12}) when $\alpha_1\neq0$ and must satisfy the constraints (\ref{eq:4.6}) and (\ref{eq:4.12}), namely
\begin{eqe}\label{eq:A.3}
\lambda^0=\pa{\delta|\vec{\lambda}^0|-\vec{v}\cdot\vec{\lambda}^0,\vec{\lambda}^0},\qquad \lambda^1=\pa{\epsilon|\vec{\lambda}^1|\pa{\frac{\kappa p}{\rho}}^{1/2}-\vec{v}\cdot\vec{\lambda}^1,\vec{\lambda}^1}.
\end{eqe}%
For the sake of simplicity, we set $e^{\Phi}=|\vec{\lambda}|e^\phi$, $f=|\vec{\lambda}|^{-1}\lambda_0$, $\vec{h}=|\vec{\lambda}|^{-1}\vec{\lambda}$ and $\tilde{\mu}=|\vec{\lambda}|^{-2}|\vec{\lambda^1}|\mu$, so that the solution (\ref{eq:A.2}) for $\lambda^0$ and $\lambda^1$ can be written in the form
\begin{aleq}\label{eq:A.4}
&\lambda_0^0=e^{\Phi(r)}f,\qquad \vec{\lambda}^0=e^{\Phi(r)}\vec{h}(r^1),\quad |\vec{h}(r^1)|=1,\\
&\lambda^1_0=\mu|\vec{\lambda}^1|\pa{\frac{\del \Phi(r)}{\del r^1}f(r^1)+\dot{f}(r^1)},\qquad \frac{\vec{\lambda}^1}{|\vec{\lambda}^1|}=\mu\pa{\frac{\del \Phi}{\del r^1}\vec{h}(r^1)+\dot{\vec{h}}(r^1)}.
\end{aleq}%
In this notation the solution for the Riemann invariants takes the form
\begin{aleq}\label{eq:A.5}
&f(r^1)t+\vec{h}(r^1)\cdot\vec{x}=\Psi(r^1)+\int_0^{r^0}e^{-\Phi(\xi,r^1)}d\xi,\\
&\dot{f}(r^1)t+\dot{\vec{h}}(r^1)\cdot\vec{x}=\dot{\Psi}(r^1)+\int_0^{r^0}\frac{\del \Phi(\xi,r^1)}{\del r^1}e^{-\Phi(\xi,r^1)}d\xi,
\end{aleq}%
where $\psi$ is an arbitrary function of $r^1$. Comparing equations (\ref{eq:A.1}.1) and (\ref{eq:A.1}.2) and repeating the procedure with equations (\ref{eq:A.1}.4) and (\ref{eq:A.1}.5), we see that the pressure and the density are related through the PDEs system
$$\frac{\del p}{\del r^i}=\frac{\kappa p}{\rho}\frac{\del \rho}{\del r^i},\quad i=1,2,$$%
which is easily solved to obtain
$$
p=A \rho^\kappa,
$$
where $A$ is a constant. Assuming that $\vec{h}\times \vec{\Omega}\neq0$, the set of vectors $\ac{\vec{h},\vec{\Omega},\vec{h}\times\vec{\Omega}}$ forms an orthonormal basis. So, we denote
\begin{eqe}\label{eq:notation}
\vec{v}=v_1\vec{h}+v_2\vec{\Omega}+v_3\vec{h}\times\vec{\Omega},\qquad \vec{g}=g_1\vec{h}+g_2\vec{\Omega}+g_3\vec{h}\times\vec{\Omega},
\end{eqe}%
where $v_i=v_i(r^0,r^1)$ and $g_i=g_i(r^1)$, $i=1,2,3$. Substituting the notation (\ref{eq:notation}) into equations (\ref{eq:A.1}.2), (\ref{eq:A.1}.3), and using (\ref{eq:A.3}) and the hypotheses $\vec{\lambda}^0\cdot\vec{\Omega}=\vec{h}\cdot\vec{\Omega}=0$, we obtain the system of PDEs
\begin{aleq}\label{eq:A.7}
&(1)\quad \frac{\del \rho}{\del r^0}=-\rho e^{-\Phi}\pa{\frac{g_1-v_3}{(f+v_1)^2-\kappa A \rho^{\kappa-1}}},\\
&(2)\quad \frac{\del v_1}{\del r^0}=-\rho e^{-\Phi}\pa{\frac{(g_1-v_3)(f+v_1)}{(f+v_1)^2-\kappa A \rho^{\kappa-1}}},\\
&(3)\quad \frac{\del v_2}{\del r^0}=e^{-\Phi}\frac{g_2}{f+v_1},\\
&(4)\quad \frac{\del v_3}{\del r^0}=e^{-\Phi}\pa{\frac{g_3-f}{f+v_1}+1}.
\end{aleq}%
It should be noted that since $\vec{\Omega}\cdot\vec{h}=0$ it follows that $\vec{\Omega}\cdot\vec{g}=g_2$ and $g_2$ is constant. Comparing equations (\ref{eq:A.7}.1) and (\ref{eq:A.7}.2), we find that $\rho^{-1}\del\rho/\del r^0=-(f+v_1)^{-1}\del v_1/\del r^0$ which leads to
$$
\rho=\frac{S(r^1)}{f(r^1)+v_1(r^0,r^1)}.
$$
By integrating of the equations (\ref{eq:A.7}.3) and (\ref{eq:A.7}.4), we obtain
\begin{aleq}\label{eq:A.9}
&(1)\quad v_2=g_2\int_0^{r^0}\frac{e^{-\Phi(\xi,r^1)}}{f(r^1)+v_1(\xi,r^1)}d\xi+V_2(r^1),\\
&(2)\quad v_3=(g_3-f)\int_0^{r^0}\frac{e^{-\Phi(\xi,r^1)}}{f(r^1)+v_1(\xi,r^1)}d\xi+\int_0^{r^0}e^{-\Phi(\xi,r^1)}d\xi+V_3(r^1),
\end{aleq}%
respectively, where $V_2(r^1)$ and $V_3(r^1)$ are functions of integration. The substitution of (\ref{eq:A.9}) into equation (\ref{eq:A.7}.2) gives
\begin{aleq*}
&\pa{v_1(r^0,r^1)+f(r^1)-\kappa A\frac{S(r^1)^{\kappa-1}}{(v_1(r^0,r^1)+f)^{\kappa}}}\frac{\del v_1}{\del r^0}=e^{-\Phi}\bigg(g_1(r^1)+V_3(r^1)\\
&\int_0^{r^0}e^{-\Phi(\xi)}d\xi-(g_3(r^1)-f(r^1))\int_0^{r^0}\frac{e^{-\Phi(\xi)}}{v_1(r^0,^1)+f(r^1)}d\xi\bigg).
\end{aleq*}%
Eliminating the function $\gamma_\rho$ from equation (\ref{eq:A.1}.6) using equation (\ref{eq:A.5}), we find
\begin{aleq*}
\frac{\vec{\lambda}^1}{|\vec{\lambda}^1|}=&-\epsilon(\kappa A)^{1/2}\rho^{-\frac{\kappa-1}{2}+1}\pa{\frac{\del \rho}{\del r^1}}^{-1}\bigg(\pa{\frac{\del v_1}{\del r^1}\vec{h}-\epsilon_2 |\dot{\vec{h}}|v_3}\vec{h}\\
&+\frac{\del v_2}{\del r^1}\vec{\Omega}+\pa{\frac{\del v_3}{\del r^1}+\epsilon_2|\dot{\vec{h}}|v_1}\vec{h}\times\vec{\Omega}\bigg),
\end{aleq*}%
where we used the fact that $\dot{\vec{h}}=\epsilon_2|\dot{\vec{h}}|(\vec{h}\times\vec{\Omega})$ as a consequence of $\vec{h}\cdot\vec{\Omega}=0$, $|\vec{h}|=1$. Next, we introduce $\lambda^0$ and $\lambda^1$ given by (\ref{eq:A.4}) into the constraints (\ref{eq:A.3}) to obtain
\begin{aleq}\label{eq:A.12}
&(1)\qquad &&\epsilon\pa{\frac{\kappa p}{\rho}}^{1/2}-\vec{v}\cdot\frac{\vec{\lambda}^1}{|\vec{\lambda}^1|}=\mu\pa{\frac{\del \Phi}{\del r^1}f+\dot{f}},\\
&(2) &&\frac{\vec{\lambda}^1}{|\vec{\lambda}^1|}=\mu\pa{\frac{\del \Phi}{\del r^1}\vec{h}=\epsilon_2|\dot{\vec{h}}|(\vec{h}\times \vec{\Omega})}.
\end{aleq}%
Combining equations (\ref{eq:A.12}.1) and (\ref{eq:A.12}.2) yields
\begin{eqe}\label{eq:A.13}
\kappa A\rho^{\kappa-2}\frac{\del \rho}{\del r^1}l^{-1}-\pa{\epsilon_2|\dot{\vec{h}}|v_3+\dot{f}}=\frac{\del \Phi}{\del r^1}\pa{f+v_1},
\end{eqe}%
and comparing equations (\ref{eq:A.12}.1) and (\ref{eq:A.12}.2) componentwise, we find
\begin{aleq}\label{eq:A.14}
&(1)\quad  &&\nu\frac{\del \Phi}{\del r^1}=\frac{\del v_1}{\del r^1}-\epsilon_2|\dot{\vec{h}}|v_3,\\
&(2) &&\frac{\del v_2}{\del r^1}=0,\\
&(3) &&\nu\epsilon_2|\dot{\vec{h}}|=\frac{\del v_3}{\del r^1}+\epsilon_2 |\dot{\vec{h}}|v_1,\\
&(4) &&\kappa A\rho^{\kappa-2}\frac{\del \rho}{\del r^1}\nu^{-1}+\epsilon_2|\dot{\vec{h}}|v_3+\dot{f}+\frac{\del \Phi}{\del r^1}(f+v_1).
\end{aleq}%
Since $|\vec{\lambda}^1|^{-1}|\vec{\lambda^1}|$ is a unit vector, the following relation holds:
\begin{eqe}\label{eq:A.15}
\pa{\pa{\frac{\del v_1}{\del r^1}-\epsilon_2|\dot{\vec{h}}|v_3}^2+\pa{\frac{\del v_3}{\del r^1}+\epsilon_2|\dot{\vec{h}}|v_1}^2}^{1/2}=-\epsilon\epsilon_4\sqrt{\kappa A}\rho^{\frac{\kappa-1}{2}}\frac{\del \rho}{\del r^1}.
\end{eqe}%
Taking the derivatives with respect to $r^1$ of each side of equation (\ref{eq:A.7}.3) and taking into account the condition (\ref{eq:A.13}.2), we obtain
$$g_2\frac{\del}{\del r^1}\pa{\frac{e^{-\Phi}}{f+v_1}}=0.$$%
If $g_2=0$, then clearly $v_2=V_2$ is a constant. If $g_2\neq0$, it is easy to see from (\ref{eq:A.9}.2) that $\dot{V}_2=0$, so $V_2$ is a constant. For the rest of this example we suppose that $|\dot{\vec{h}}|\neq0\neq\del\Phi/\del r^1$. It has been shown that the other subcases do not admit any solution. As an illustration, we consider only the case when $g_2\neq0$. Using equation (\ref{eq:A.7}.3), the solution for $v_3$ given by (\ref{eq:A.9}.2) can be rewritten in the form
\begin{eqe}\label{eq:A.16}
v_3=g_2^{-1}v_2(r^0)+g_2^{-1}\int_0^{r^0}\dot{v}_2(\xi)(f(r^1)+v_1(\xi,r^1))d\xi+V_3(r^1).
\end{eqe}%
Since $v_2$ is a function of $r^0$ only, we can take the derivative of the equation (\ref{eq:A.7}.3) with respect to $r^1$ to determine the derivative $\del \Phi/\del r^1$. We obtain
\begin{eqe}\label{eq:A.17}
\frac{\del \Phi}{\del r^1}=-\frac{1}{v_1+f}\pa{\frac{\del v_1}{\del r^1}+\dot{f}}.
\end{eqe}%
Using equations (\ref{eq:A.17}) and (\ref{eq:A.14}.1) to eliminate the quantities $\del \Phi/\del r^1$ and $\mu$ respectively from equations (\ref{eq:A.14}.3) and (\ref{eq:A.14}.4), we obtain
\begin{aleq}\label{eq:A.18}
&(1)\quad &&\epsilon_2|\dot{\vec{h}}|\pa{v_1+f}\pa{\frac{\del v_1}{\del r^1}-\epsilon_2|\dot{\vec{h}}|v_3}+\pa{\frac{\del v_1}{\del r^1}+\dot{f}}\pa{\frac{\del v_3}{\del r^1}+\epsilon_2|\dot{\vec{h}}|v_1}=0,\\
&(2) &&\pa{\frac{\del v_1}{\del r^1}-\epsilon_2|\dot{\vec{h}}|v_3}\pa{\frac{\del v_3}{\del r^1}+\epsilon_2|\dot{\vec{h}}|v_1}=\epsilon_2\kappa A|\dot{\vec{h}}|\rho^{\kappa-2}\frac{\del \rho}{\del r^1}.
\end{aleq}%
Substituting $\rho=(v_1+f)^{-1}S$ into equations (\ref{eq:A.18}) and (\ref{eq:A.15}), we find
\begin{eqe}\label{eq:A.19}
\pa{\frac{\del v_1}{\del r^1}-\epsilon_2|\dot{\vec{h}}|v_3}\pa{\frac{\del v_3}{\del r^1}+\epsilon_2|\dot{\vec{h}}|v_1}=\epsilon_2\kappa A|\dot{\vec{h}}|\pa{\frac{S}{v_1+f}}^{\kappa-2}\frac{\del}{\del r^1}\pa{\frac{S}{v_1+f}}
\end{eqe}%
and
\begin{eqe}\label{eq:A.20}
\pa{\pa{\frac{\del v_1}{\del r^1}-\epsilon_2|\dot{\vec{h}}|v_3}^2+\pa{\frac{\del v_3}{\del r^1}+\epsilon_2|\dot{\vec{h}}|v_1}^2}^{1/2}=-\epsilon\epsilon_4\sqrt{\kappa A}\rho^{\frac{\kappa-1}{2}}\frac{\del }{\del r^1}\pa{\frac{S}{v_1+f}},
\end{eqe}%
respectively. Equation (\ref{eq:A.7}.1) becomes
\begin{eqe}\label{eq:A.21}
\frac{\del }{\del r^0}\pa{\frac{s}{v_1+f}}=-S\dot{v}_2\pa{\frac{g_1-v_3}{(f+v_1)^2-\kappa A \rho^{\kappa-1}}}.
\end{eqe}%
In summary, the solutions are defined by
\begin{gaeq}\label{eq:A.22}
\rho=\frac{S(r^1)}{v_1(r^0,r^1)+f(r^1)},\qquad p=A \rho^{\kappa},\\
\vec{v}=v_1(r^0,r^1)\vec{h}(r^1)+v_2(r^0)\vec{\Omega}+v_3(r^0,r^1)\vec{h}(r^1)\times\vec{\Omega},\\
\end{gaeq}%
where $v_3(r^0,r^1)$ is given by (\ref{eq:A.16}) and the functions $v_1(r^0,r^1)$, $v_2(r^0)$, $S(r^1)$, $f(r^1)$ and $|\dot{\vec{h}}|$ can be chosen in order to satisfy the conditions (\ref{eq:A.18}.1) and (\ref{eq:A.19}-\ref{eq:A.21}). The invariants $r^0$ and $r^1$ are defined implicitly by equation (\ref{eq:A.5}). The results of the superposition of simple waves on simple states written in terms of Riemann invariants can be classified according to the definition given in \cite{Jeffrey:1976}. The propagation of an acoustic simple wave on a hydrodynamics simple state, given by expressions (\ref{eq:A.22}), is strictly non-linear. Indeed, for all $i,j\in\ac{0,1}$ such that $i\neq j$, the relation $\del \lambda^i/\del r^j\neq 0$ holds. Hence the change of the direction of propagation of the acoustic simple wave is genuinely influenced by the hydrodynamic state.
\section{Future outlook}
In this paper we have shown that the propagation of single simple waves described by the inhomogeneous multidimensional Euler equations (\ref{eq:euler}) in (3+1) dimensions does occur and can be expressed in terms of Riemann invariants. The results presented herein are a very important class of solutions, since they are ubiquitous for the Euler equations and constitute their elementary rank-2 solutions. These solutions are the building blocks for constructing more general types of solutions describing superpositions of many waves (\textit{i.e.} rank $k>2$), which are more interesting from the physical point of view. Until now, the only way to approach this task was through the method of characteristics, which relies on treating Riemann invariants as new dependent variables (which remain constant along appropriate characteristic curves of the initial system). This leads to the reduction of the dimensionality of the problem. The determination of necessary and sufficient conditions for the existence of Riemann $k$-waves in multidimensional inhomogeneous systems of PDEs allows us to look at higher rank solutions (\textit{i.e.} greater than 2) \cite{Grundland:1974:b,Peradzynski:1971:planewave}. As it was shown in \cite{Grundland:1974:b} this type of solution depends on some arbitrary functions of one variable. The criteria for determining the elastic or nonelastic character of the nonlinear superpositions of waves expressed in terms of Riemann invariants are particularly useful in physical applications. However, the method of characteristics, like all other techniques for solving PDEs, has its limitations. It is well known \cite{Friedrich:1948,Rozdestvenski:1983} that solutions of the Euler equations, even with arbitrarily smooth initial data, usually cannot be extended indefinitely in time. After a certain finite time period they blow up. The first derivatives of a solution become unbounded after some time. Solutions of the initial value problem do not exist and the gradient catastrophe can occur. Non-continuous solutions occur in the form of shock waves \cite{Friedrich:1948,Rozdestvenski:1967,Whitham:1974}. In the next stage of this research, an analysis of the asymptotic behavior of such solutions and the patterns of wave superpositions (\textit{e.g.} scattering and nonscattering) will be conducted. Among other possibilities, the conditions leading to the phenomena of the gradient catastrophe and the appearance of shock waves will be investigated. This task will be undertaken in our future work.
\section*{Acknowledgements}This project was completed during A.M.G.'s one month visit to \'Ecole Normale Sup\'erieure de Cachan and he would like to thank the Centre de Math\'ematiques et de leurs Applications (CMLA) for their kind invitation. A.M.G. has been supported by the Fondation Math\'ematique Jacques Hadamard (FMJH). This work was supported by ANR-11-LABX-0056-LMH, LabEx LMH.
\paragraph{}A.M.G. would also like to thank the Ph.D. school in Physics of Roma Tre University for one month's financial support and the Departement of Mathematics and Physics for their hospitality.
\paragraph{}A.M.G.'s work was supported by a research grant of the Natural Sciences and Engineering Research Council of Canada (NSERC). This project was completed during V.L.'s visit to the Centre de Recherches Mathématiques (CRM) of the Université de Montréal and he would like to thank the CRM for their kind invitation and hospitality.
\bibliographystyle{alpha}

\end{document}